\documentclass[a4paper,times,doublespace,11pt,onecolumn]{article}

\usepackage{graphicx}% Include figure files
\usepackage{dcolumn}% Align table columns on decimal point
\usepackage{bm}% bold math

\usepackage[utf8]{inputenc}
\usepackage[T1,T5]{fontenc}

\usepackage{etoolbox}
\usepackage{dsfont}
\usepackage{hhline}
\usepackage{enumitem}
\usepackage{algorithm}
\usepackage{url}
\usepackage{amsmath}
\usepackage{amsthm}
\usepackage{amssymb}
\usepackage{upgreek} 
\usepackage{booktabs}

\newtheorem{theorem}{Theorem}

\newtheorem{definition}{Definition}
\theoremstyle{remark}
\newtheorem{remark}{Remark}

\providecommand{\norm}[1]{\lVert#1\rVert}

\newcommand{\pr}{\text{proj}}

\providecommand{\norm}[1]{\lVert#1\rVert}

\def\bbR{\mathbb{R}}
\def\bbNz{\mathbb{N}_0}

\def\bc{{\bm c}}

\def\bh{{\bm h}}

\def\bxi{\bm \xi}
\def\bkappa{\bm \kappa}
\def\bw{\bm w}

\def\bu{{\bm u}}
\def\bv{{\bm v}}
\def\bx{{\bm x}}
\def\by{{\bm y}}

\def\bA{{\bm A}}
\def\bB{{\bm B}}
\def\bC{{\bm C}}
\def\bD{{\bm D}}
\def\bE{{\bm E}}
\def\bF{{\bm F}}
\def\bG{{G}}
\def\bI{{\bm I}}
\def\bJ{\bm J}

\def\bH{{\bm H}}

\def\bR{{\bm R}}
\def\bS{{\bm S}}

\def\bU{{\bm U}}
\def\bV{{\bm V}}
\def\bY{{\bm Y}}

\def\bX{{\bm X}}

\def\bW{{\bm W}}

\def\bnu{{\bm \nu}}

\def\TNFA{TNF-$\alpha$ }
\def\NFKB{NF-$\kappa$B }

\DeclareMathOperator*{\argminB}{argmin}   % Jan Hlavacek

\def\k0{k_*}

\def\TNFA{{TNF$\alpha$} }
\def\NFKB{NF-$\kappa$B }

\def\vec1{\text{vec}}

\def\NFKB{NF-$\kappa$B }
\def\S0{\bS_0}

\def\cL{\mathcal{L}}

\def\bzero{{\bm 0}}

\usepackage{authblk}

\title{Simulating stochastic population dynamics: The Linear Noise Approximation can capture non-linear phenomena}% Force line breaks with \\
\author[a]{Frederick Truman-Williams}
\author[b,1]{Giorgos Minas}

\affil[a]{Mathematical Institute, University of Oxford, Oxford, OX2 6GG, United Kingdom}
\affil[b]{School of Mathematics and Statistics, University of St Andrews, St Andrews, KY16 9SS, Scotland, United Kingdom\footnote{To whom correspondence should be addressed. E-mail: gm256@st-andrews.ac.uk}}

\date{}

\begin{document}

\maketitle

\begin{abstract}
Population dynamics in fields such as molecular biology, epidemiology, and ecology exhibit highly stochastic and non-linear behaviour. In gene regulatory systems in particular, oscillations and multi-stability are especially common.
Despite this, none of the currently available stochastic models for population dynamics are 
both accurate and computationally efficient for long-term predictions. A prominent model in this field, the Linear Noise Approximation (LNA), is computationally efficient for tasks such as simulation, sensitivity analysis, and parameter estimation; however, it is only accurate for linear systems and short-time predictions. Other models may achieve greater accuracy across a broader range of systems, but they sacrifice computational efficiency and analytical tractability. This paper demonstrates that, with specific modifications, the LNA can accurately capture non-linear dynamics in population processes. We introduce a new framework based on centre manifold theory, a classical concept from non-linear dynamical systems. This approach enables the identification of simple, system-specific modifications to the LNA, tailored to classes of qualitatively similar non-linear dynamical systems. With these modifications, the LNA can achieve accurate long-term simulations without compromising computational efficiency. We apply our methodology to classes of oscillatory and bi-stable systems, and present multiple examples from molecular population dynamics that demonstrate accurate long-term simulations alongside significant improvements in computational efficiency.
\end{abstract}

\section{Introduction}
In molecular biology, ecology, epidemiology and other fields, there are many non-linear dynamical phenomena such as oscillations and multi-stability.
Examples include circadian rhythms \cite{drosmodel}, embryonic development oscillations \cite{Marinopoulou2021}, cell signalling oscillations \cite{ashall}, multi-stability in the cell cycle \cite{Novak1993,Pomerening2005,Angeli2004}, cell differentiation \cite{Corson2012,Goldbeter2007}, and apoptosis \cite{Arancibia}, as well as the central carbon metabolism in E. coli \cite{IshiiNobuyoshi2007MHAM}, predator-prey oscillations in ecology \cite{froda07}, and epidemic oscillations \cite{Weitz2020}.
These non-linear dynamics arise from feed-forward loops \cite{Mangan2003}, negative feedback loops \cite{Mangan2003,Rosenfeld2002} and other mechanisms governed by networks of interacting species.
For example, gene expression oscillations are often generated when a gene directly suppresses its own expression or activates the expression of its suppressors. 
In infectious disease outbreaks, compartments of the population, such as infected and susceptible individuals, interact with each other and phenomena like awareness and fatigue can generate oscillatory behaviours \cite{Weitz2020}.
The effects of stochasticity in interactions between different populations—arising from molecular diffusion or, more generally, from the complex environments in which these interactions occur—are significant, particularly for small populations. Yet modelling this stochasticity using fast, accurate, and scalable methods remains a challenging open problem.

Networks of stochastic interactions between different populations are often called
reaction networks.
The term ‘reaction’ refers not only to biochemical reactions, but to any event that directly changes the population of one or more species or groups.
The stochastic dynamics of reaction networks can be described, under certain assumptions, by a Markov process
$\{\bY(t) = (Y_1(t),\dots,Y_n(t))^\top:   t\geq 0\}$, 
describing the evolution of $n$ different populations
over time.
The so-called (chemical) master equation \cite{Gill76}, which describes the evolution of the probability distribution of the species populations over time, is analytically solvable only in rare cases, severely limiting its tractability.
The stochastic simulation algorithm (SSA), also known as the Gillespie algorithm \cite{Gill77}, 
produces exact simulations of the Markov process $\{\bY(t):t\geq 0\}$ but the SSA 
is far too slow to be used in most practical applications
since it simulates every single reaction event.

At present, there are several methods available for approximating the master equation, which aim to accelerate simulations. 
Notable examples include tau leaping algorithms \cite{Gill2001,Gill2003} and the numerical integration of the chemical Langevin equations \cite{Gill2000}.
Both approaches significantly increase simulation speed; however, they leave no scope for the analytical study of the probability distributions of the populations and can be extremely slow for sensitivity analysis and parameter estimation. 
Other recent methods \cite{Cao2018,Jia2024} focus on simplifying protein-binding reactions, which frequently appear in molecular biology networks.

Much work has followed from van Kampen's system size expansion \cite{Kampen1992}, which uses a system size parameter $\Omega$ (like cell volume in molecular biology, maximum population size in ecology or epidemiology) to derive approximations of the 
master equation.  
Under certain conditions, the deterministic process $\{\bx(t)=(x_1(t),\dots,x_n(t))^\top :   t\geq 0\}$ which solves the macroscopic reaction-rate equations has been shown to be the law of large numbers limit, as $\Omega\to\infty$, of the Markov process governed by the master equation
\cite{kurtz_1971,kurtz1981approximation}.
The Linear Noise Approximation (LNA), derived using the system size expansion method, is a stochastic model that  
provides even more rapid simulation than the aforementioned stochastic methods as well as being the only model in this field that 
allows for analytical expressions of probability distributions of populations at arbitrary time-points.
This makes the LNA a powerful tool for sensitivity analysis and statistical inference \cite{Minas2017,MinasGiorgos2019Psaf,SwallowBen2022Bifs,MinasGiorgos2020Mift,Komorowski2009,Finkenstadt2013,Golightly2023,Schnoerr2017}.
However, the important drawback of using the LNA is its inability to approximate the long-term behaviour of non-linear population dynamics. 
Indeed, when the system dynamics are non-linear \cite{Wallace2012}, predictions of the state of the system at time $t$ given the state at an earlier time $s$, $0<s<t$, become inaccurate as $t-s$ becomes large. 

For any given system, trajectories can be examined qualitatively to identify distinct phases in the population dynamics, such as peaks and troughs. However, when considering the distribution of SSA-generated trajectories at a finite set of time points, their phases become increasingly misaligned with those predicted by the LNA.
As a result, the LNA makes incorrect predictions of the dynamics thereafter. 
In \cite{Minas2017}, it was shown that in the specific setting where the macroscopic ($\Omega\to \infty$) deterministic approximation has an attractive limit cycle solution, correcting these drifts in phase predicted by the LNA is sufficient to produce simulations that are nearly identical to those produced by the SSA.
This approach enables the construction of both fast and accurate algorithms for stochastic simulation, sensitivity analysis, and statistical inference of these
non-linear systems \cite{Minas2017}. 

The main results of this paper are the following.
Firstly, we develop a novel framework for constructing LNA-based models for non-linear dynamical systems. 
We adopt the term phase-corrected LNA (pcLNA)—originally introduced in \cite{Minas2017}—to describe this stochastic modelling approach, which evolves the system using LNA equations while periodically adjusting the phase to mitigate phase drift.
Fundamental to the pcLNA approach is a mapping $\bG$ that computes the phase of any stochastic state,
allowing for phase drift to be corrected. 
We show that, for any parameter-dependent system whose macroscopic reaction-rate equation admits a non-hyperbolic equilibrium for some parameter value, tools from dynamical systems theory can be used to explicitly construct the phase-correcting map $\bG$.
Notably, by combining results from \cite{Takens1983,Takens1971} with Itô’s Lemma \cite{LeGallsdebook}, we derive a novel decomposition of the LNA stochastic process into coordinates exhibiting non-linear, persistent dynamics and coordinates with transient dynamics. This decomposition suggests that a common phase definition can be used for classes of stochastic systems with qualitatively similar behaviour.
Using these findings, we provide a step-by-step procedure for defining a phase-correcting map and a general algorithm to be used for stochastic simulation.

Secondly, we use this framework to provide readily applicable algorithms that can accurately simulate long-term stochastic dynamics for arguably the two most abundant non-linear phenomena: oscillations and bi-stability.
We demonstrate this on several systems including the oscillatory systems in 
\cite{LEFEVER1971267,WILHELM1995,ashall}, 
and bi-stable systems in \cite{Gardner2000,Angeli2004,Goldbeter2007}. 
Here the pcLNA model offers accurate approximations of the master equation for long-term simulation whilst substantially reducing the simulation time by several orders of magnitude.
We also propose a fast method to predict the speed of SSA simulations, and therefore assess whether a faster simulation method is necessary.

We emphasize that with further development on defining explicit phase-correcting maps $\bG$, 
the pcLNA framework introduced here could be used to simulate other classes of non-linear dynamical systems with computational efficiency and accuracy.
We also note that a key implication of the speed and accuracy of the pcLNA is that it enables the approach to be used for computationally intensive tasks such as sensitivity analysis, inference, and prediction for large, non-linear systems.

While the approach developed in \cite{Minas2017}
demonstrated that phase correction significantly improves the long-term accuracy of the LNA in oscillatory systems with stable limit cycles, it was limited to this specific dynamics. In contrast, the framework introduced in this paper generalises the concept of phase correction by introducing centre manifold theory to define and correct phase drifts to a broad class of systems only assumed to have a non-hyperbolic equilibrium, which includes the abundant classes of oscillatory and bi-stable systems.

We begin in Section \ref{prelims} with a review of stochastic population dynamics evolving through reaction networks (Section~\ref{sec:reactionnetworks}) and the system size expansion (Section~\ref{sec:systemsize}), and then introduce the Linear Noise Approximation (LNA) in Section~\ref{sec:LNA}.
We conclude the section by discussing the LNA's loss of accuracy in non-linear systems. 
In Section~\ref{sec:pcLNA}, we introduce the phase-corrected Linear Noise Approximation framework. 
To understand the LNA's failure we define the notion of phase in the context of stochastic reaction networks (Section~\ref{sec:phase}) before presenting the general phase-corrected LNA (pcLNA) simulation algorithm in Section~\ref{pcLNAsect}.
We follow this with a discussion on how to define the phase-correcting map in Section~\ref{sec:phaseCorrectionMap}. In Section~\ref{sec:hopf}, we apply our framework to reaction networks whose macroscopic reaction rate equation undergoes a Hopf bifurcation and develop the pcLNA algorithm specific to these systems. Section~\ref{sec:bi-stable} focuses on bi-stable systems and constructs the corresponding pcLNA algorithm. We validate our framework through numerical investigations in Section~\ref{sec:numerical_investig}, and conclude with a discussion of our findings and future directions in Section~\ref{discussion}. Additional numerical investigations demonstrating the accuracy of the pcLNA modelling techniques on other reaction networks, as well as supporting technical results and derivations, are provided in the Supplementary Material (SM).

\begin{figure}[h]
    \centering
    \includegraphics[width=0.5\textwidth]{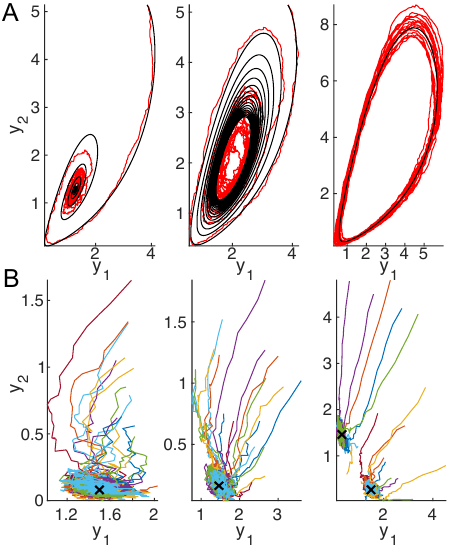}
    \caption{\label{fig:dynamics} A. Phase portraits of solutions of the RRE in \eqref{eq:bifode1} (black) and an SSA simulation (red) for the system in \cite{WILHELM1995}. 
    B. Phase portraits of 20 SSA simulations (color) and the equilibria of the RRE in \eqref{eq:bifode1} (crosses) for the system in \cite{Gardner2000}. The qualitative changes from left to right in the figure are due to changes in the bifurcation parameter value.
    }
\end{figure}

\section{Population dynamics}
\label{prelims}

We begin, in Sections \ref{sec:reactionnetworks}-\ref{sec:LNA}, by reviewing the mathematical formulation of population dynamics evolving through reaction networks. 
Our discussion is restricted to the concepts needed for the development of our methodology; for a more comprehensive review, see \cite{Anderson2011,Masuda2023,Wilkinson2012}.
We describe this formulation within the context of molecular biology, although analogous terminology can be applied in other fields such as epidemiology and ecology (see, for instance, \cite{Golightly2023,Wilkinson2012} and Figure \ref{fig:LNA}). 

\subsection{Reaction networks}\label{sec:reactionnetworks}

A system of $n$ different molecular species, $M_1,M_2,\dots,M_n$ has state vector
\begin{equation}
    \bY(t) = (Y_1(t), \ldots ,Y_n(t))^\top
\end{equation}
where 
$Y_i(t)$, $i \in\{ 1, \ldots , n\}$, denotes the number of molecules (i.e.\ the population) of species $M_i$ at time $t$ and $\top$ denotes the transpose. These molecules undergo reactions $R_j$, $j\in\{1,\dots,r\}$, such as transcription, translation, degradation and translocation, which change the number of molecules of each
species,
\begin{equation}\label{eq:reaction}
    k_{1j} M_1 + \dots + k_{nj} M_n \stackrel{c_j}{\longrightarrow} k_{1j}' M_1 + \dots + k_{nj}' M_n.
\end{equation}
Here, the non-negative integer coefficients $k_{ij}$ and $k_{ij}^\prime$ 
denote the number of molecules of species $M_i$ involved as reactants and products, respectively, in reaction $R_j$. 
If $R_j$ occurs at time $t\geq 0$, $\bY(t)$ jumps to a new state
$\bY(t) + \bnu_j$, where the integer transition vectors $\bnu_j = (k_{1j}'-k_{1j},\dots,k_{nj}'-k_{nj})^\top$ are often called stoichiometric.
Under the assumption that these reactions occur in a well-mixed solution \cite{Gill77}, we can derive the propensity functions $\pi_j:\bbR^n\to[0,\infty)$ for each $j\in\{1,\dots,r\}$ with the probability that one $R_j$ reaction occurs in the next infinitesimal time interval $[t,t+dt)$
\begin{equation}\label{eq:prop}
    P(\bY(t+dt)-\bY(t) = \bnu_j) = \pi_j(\bY(t))dt + o(dt).
\end{equation}
The propensity functions $\pi_j$ along with the stoichiometric vector $\bnu_j$, $j=1,\dots,r$, completely specify the evolution of the reaction network.

The general form of the propensity function of the reaction in \eqref{eq:reaction}, when $\bY(t)=\by=(y_1,\dots,y_n)^\top$ is $\pi_j(\by) = c_j\prod_i \binom{y_i}{k_{ij}}\footnote{$\binom{n}{k}$ is the binomial coefficient of $n$ choose $k$.}$. 
Uni-molecular ($M_i\stackrel{c_j}{\longrightarrow}\; \cdot$) and bi-molecular ($M_i + M_{i'}\stackrel{c_j}{\longrightarrow} \; \cdot$) reactions, with propensities $c_jy_i$ and $c_jy_iy_{i'}$ respectively, are the most common types (see \cite{WaageP1986Sca}).
Here $c_j>0$ is the reaction rate constant corresponding to reaction $R_j$. The propensity functions are not restricted to the above general form. For instance, for enzymatic reactions involving species $M_i$ the propensity has the Michaelis-Menten form \cite{JohnsonKennethA2011TOMC}, $c_2y_i/(c_1+y_i)$, while the Hill form \cite{Hill1910}, $c_2y_i^h/(c_1^h+y_i^h)$, is also often used to reflect cooperative binding.
For a discussion on the latter forms of propensities see \cite{Cao2018}. 

These conditions give rise to a continuous-time, discrete state-space Markov stochastic process $\{\bY(t) :   t\geq 0\}$ with state-dependent propensity function. 
The Kolmogorov forward equation \cite{Gill92} describing the evolution of the probability distribution, $P(\by,t) := P(\bY(t)=\by \mid \bY(0)=\by_0 )$, of the stochastic process $\{\bY(t) : t\geq 0\}$,
\begin{equation}\label{eq:master}
	\frac{\partial P(\by,t)}{\partial t} = \sum_{j=1}^r \pi_j(\by-\bnu_j) P(\by-\bnu_j,t) - \sum_{j=1}^r \pi_j(\by) P(\by,t),	
\end{equation}
is often referred to as the (chemical) master equation. 
Despite the equation only being analytically tractable for few cases, exact trajectories can be generated using Gillespie's stochastic simulation algorithm (SSA) \cite{Gill77}. We provide details of this in SM (Section \ref{app:SSA}). 

Equivalently, the process can be represented via the random time change representation (RTC) \cite{Anderson2011}
\begin{equation}\label{eq:rtc}
	\bY(t) = \bY(0) + \sum_{j=1}^r \bnu_j N_j \left( \int_0^t \pi_j(\bY(s))ds  \right),
\end{equation} 
where, for each $j\in\{1,\dots,r\}$, $N_j$ is an independent, unit-rate Poisson process corresponding to reaction $R_j$.
For a given trajectory $\by(s)$, $s\in[0,t)$, the random variables $N_j \left( \int_0^t \pi_j(\by(s))ds  \right)$ are independent and have Poisson distribution with mean $\int_0^t \pi_j(\by(s))ds$.

\paragraph{Example (SIR model)} 
The SIR model used to describe population dynamics in infectious disease outbreaks is an example of a reaction network. 
We formulate it here to provide a concrete instance of the general framework introduced above (see also Figure \ref{fig:LNA}(A-B) for a summary).
In the SIR network, there are three populations, say $M_1,M_2,M_3$, corresponding to the Susceptible (S), Infected (I), and Recovered (R) individuals. 
Two reactions can occur—infection ($R_1$) and recovery ($R_2$)—which can be represented as
$$ S + I \xrightarrow{c_1} 2I , \qquad I \xrightarrow{c_2} R, $$
where $c_1$ and $c_2$ are the reaction rate constants for the two reactions. 
The state vector $\bY=(Y_1,Y_2,Y_3)^\top$ represents the three populations $M_1$ (S), $M_2$ (I), and $M_3$ (R). 
The propensity functions are
$$ \pi_1(\bY) = c_1Y_1Y_2, \qquad \pi_2(\bY) = c_2Y_2 $$
with corresponding stoichiometry vectors $\bnu_1=(-1,+1,0)^\top$ and $\bnu_2=(0,-1,+1)^\top$. 
This can be used to formulate Equations \eqref{eq:master} and \eqref{eq:rtc} for this system.

\subsection{System Size Expansion}\label{sec:systemsize}

Following the work of \cite{kurtz_1971} and \cite{Kampen1992}, it is common in seeking methods to approximate the process $\{\bY(t) :   t\geq 0\}$, to introduce a system size parameter $\Omega$.
Through setting $\bX(t) := \bY(t)/\Omega$, one can study the dependence of stochastic fluctuations upon system size. 
It is sufficient to assume that the rates $\pi_j(\bY(t))$ depend upon $\Omega$ as $\pi_j(\bY(t)) := \Omega \rho_j(\bY(t)/\Omega)$ where $\rho_j:\bbR^n\to[0,\infty)$ are called  macroscopic rates of the system and are smooth. This is a fairly relaxed condition that can be further relaxed to more general relations between the two functions \cite{Kampen1992,kurtz1981approximation}.

Using this condition, we can re-write the infinitesimal RTC equation in \eqref{eq:rtc} in terms of $\bX(t)$ as
\begin{equation}\label{eq:masterx}
    \bX(t+dt)-\bX(t) = \sum_{j=1}^r \frac{\bnu_j}{\Omega} N_j\left(\Omega \rho_j( \bX(t)) dt \right), \quad dt>0
\end{equation}
with $\bX(0)=\bX_0$, thus providing a time evolution law for the process $\{\bX(t) :   t\geq 0\}$. If, for each $t\geq 0$, we define $\bx(t)$ as the limit in probability of $\bX(t)$ as $\Omega \to \infty$, we can use the law of large numbers\footnote{Note that for a sequence $X_1,X_2,\dots$, of independent Poisson random variables with mean $0<\lambda<\infty$, $N^{-1}\sum_{i=1}^N X_i$ converges in probability to $\lambda$, as $N\to \infty$.} (see \cite{kurtz_1971,kurtz1981approximation}) to derive
the limit of equation 
$\eqref{eq:masterx}$, as $\Omega \to \infty$, 
$$ \bx(t+dt) - \bx(t) = \sum_{j=1}^r \bnu_j   \rho_j(\bx(t)) dt, \qquad 
\bx(0)=\bx_0.$$	
Letting $dt\to 0$ we derive the macroscopic reaction rate equation (RRE), corresponding to the classical mass-action kinetics. 
A typical requirement in applied biochemical settings 
is to study the qualitative changes in the dynamics of $\{\bX(t) :   t\geq0\}$ and $\{\bx(t) :   t\geq0\}$ when one (or more) of the system parameters -- typically reaction rate constants -- are allowed to vary smoothly. Let this parameter be given by $\alpha\in\bbR$ and write the RRE as 
\begin{equation}\label{eq:bifode1}
\dot{\bx} = \frac{d\bx}{dt} =\bF(\bx,\alpha), \quad  \bF(\bx,\alpha) 
:=\sum_{j=1}^r \bnu_j \rho_j(\bx,\alpha), \qquad \bx(0)=\bx_0. 
\end{equation}
where both $\rho_j$ and $\bF$ depend smoothly on $\bx$ and $\alpha$.
The corresponding Jacobian matrix  is
\begin{equation}\label{eq:jac}
\bJ=\bJ(\bx,\alpha)=(\partial F_i/\partial x_j)_{i,j=1}^n.
\end{equation}
We will refer to the parameter $\alpha$ as the bifurcation parameter.

\paragraph{Example (SIR model, continued)} If we define a system size parameter $\Omega$ (e.g. total population) for the SIR model above, then we can apply system size expansion. The reaction rates are chosen so that $\pi_j(\bY(t)) = \Omega \rho_j(\bY(t)/\Omega)$, and hence for $\bx=\bY/\Omega$ we get
\begin{align*}
 \pi_1(\bY) &= c_1 Y_1 Y_2 = \Omega \tilde{c}_1 \,x_1 x_2 = \Omega \rho_1(\bx),&& \rho_1(\bx)=\tilde{c}_1 \,x_1 x_2,\quad \tilde{c}_1 = \Omega c_1 \\ 
 \pi_2(\bY) &= c_2 Y_2 = \Omega c_2 x_2 = \Omega\rho_2(\bx), &&  \rho_2(\bx)= c_2 x_2 
\end{align*}
which together with the stoichiometry vectors give the RRE 
\begin{align*}
  \dot{x}_1 = dx_1/dt &= -\tilde{c}_1 x_1 x_2  \\
  \dot{x}_2 = dx_2/dt &= +\tilde{c}_1 x_1 x_2 - c_2 x_2 \\
  \dot{x}_3 =  dx_3/dt &= +c_2 x_2.
\end{align*}

\subsection{The Linear Noise Approximation (LNA)}\label{sec:LNA}

To take into account the stochasticity of interactions between (molecular or other) populations, one approach is to define the LNA ansatz equation \cite{Kampen1992,kurtz_1971,kurtz1981approximation} that describes the relation between the stochastic process $\{\bX(t) :   t\geq 0\}$ and a deterministic solution $\gamma=\{\bx(t) :   t\geq 0\}$ of the RRE system in \eqref{eq:bifode1}. 
Their difference, scaled by $\Omega^{-1/2} $, is a stochastic process, $\{\bxi(t) :   t\geq 0\}$, with $\bxi(t)$ quantifying the noisy deviations away from $\bx(t)$. That is, for each $t\geq 0$,
\begin{equation}\label{eq:lnaansatzX}
    \bX(t) = \bx(t) + \Omega^{-1/2}\bxi(t).
\end{equation}
The time-evolution of $\{\bxi(t),t\geq 0\}$, for sufficiently large $\Omega$, can be described (see \cite{Minas2019a} or SM Section~\ref{app:lna} for the derivation details) by the linear stochastic differential equation (SDE) in the It\^{o} sense
\begin{equation}\label{eq:langevin}
    d\bxi = \bJ_t\bxi dt + \bE_t\, d\!\bB_t,
\end{equation}
where $\bJ_t=\bJ(\bx(t))$ is the Jacobian matrix in \eqref{eq:jac} evaluated at $\bx(t)$,  $\bB_t$ an $r$-dimensional Wiener process \cite{Wilkinson2012}, and 
\begin{equation}\label{eq:Ematrix}
\bE_t=\bE(\bx(t))=\bS\text{diag}\left(\!\sqrt{\rho_1(\bx(t))},\dots,\sqrt{\rho_r(\bx(t))}\right),
\end{equation}
the product of the stoichiometry matrix
$\bS=[\bnu_1 \cdots \bnu_r]$ and the square root of the diagonal matrix with main diagonal entries being the reaction rates $\rho_1(\bx(t)),\dots,\rho_r(\bx(t))$.
This linear SDE in \eqref{eq:langevin} has a solution that, for $0\leq s < t <\infty$, can be written as  
\begin{equation}\label{eqn:xistate}
    \bxi(t) = \bC(s,t) \bxi(s) + {\bm \eta}(s,t), \quad {\bm \eta}(s,t) \sim \text{MVN}(\bzero,\bD(s,t)),
\end{equation}
where $\bC(s,t)$ is the fundamental matrix of \eqref{eq:bifode1}, i.e.
the solution of the initial value problem
\begin{equation}\label{eq:Code}
   \dot{\bC} = \bJ_t\bC, \quad \bC(s,s) = \bI,
\end{equation}
and $\bD(s,t)$ is the symmetric, positive-definite matrix that is a solution of the initial value problem
\begin{equation}\label{eq:Vode}
    \dot{\bD} = \bJ_t\bD + \bD\bJ_t^\top + \bE_t\bE_t^\top, \quad \bD(s,s) = \mathbf{0}.
\end{equation} 
Here we write $\bI$ for the identity matrix and $\mathbf{0}$ for the zero matrix. 
We also write $\text{MVN}({\bm m},{\bm S})$ for the Multivariate Normal (Gaussian) distribution with mean ${\bm m}$, and covariance matrix ${\bm S}$.

The above representation implies that by solving the initial value problems in $\eqref{eq:bifode1}$, $\eqref{eq:Code}$ and $\eqref{eq:Vode}$, and starting with an initial condition for $\bxi(0)=\bxi_0$ drawn from an arbitrary MVN distribution, one can easily derive the MVN probability distribution of $\bxi(t)$, for any time $t>0$. In this case, we say that the LNA prediction for $\bX(t)$ was initialised at time $s=0$. 
The process $\{\bX(t) :   t\geq 0\}$ modelled by the LNA is only an approximation of the true dynamics of the reaction network $\{\bY(t)/\Omega :   t\geq 0\}$ where $\{\bY(t) :   t\geq 0\}$ evolves according to \eqref{eq:rtc}.

Therefore, when we refer to the notion of accuracy, we mean the accuracy in the distribution of $\bX(t)$ when trying to capture the distribution of $\bY(t)/\Omega$ for each $t\geq 0$, where $\{\bY(t) :   t\geq 0\}$ evolves according to \eqref{eq:rtc}.

\paragraph{Example (SIR model, continued):} To apply the LNA to the SIR reaction network, we first need to compute the Jacobian matrix by differentiation, which gives 
$$ \bJ = \left(  
\begin{array}{ccc} 
- \tilde{c}_1 x_2 & - \tilde{c}_1 x_1 & 0 \\
\tilde{c}_1 x_2 & \tilde{c}_1 x_1 - c_2  & 0 \\
0 & c_2 & 0   
\end{array} 
\right)$$ 
and the matrix defined in \eqref{eq:Ematrix}, which is
$$ \bE = \left(  
\begin{array}{cc} 
-1 & 0 \\
+1 & -1 \\
0 & +1   
\end{array} 
\right)
\left(
\begin{array}{cc} 
\sqrt{\tilde{c}_1  x_1 x_2} & 0 \\
0 & \sqrt{c_2x_2} \\
\end{array} 
\right)=
\left(  
\begin{array}{ccc} 
-\sqrt{\tilde{c}_1  x_1 x_2} & 0  \\
\sqrt{\tilde{c}_1  x_1 x_2} & -\sqrt{c_2x_2} \\
0 & \sqrt{c_2x_2}   
\end{array} 
\right). $$
These matrices depend on the solution, $\{\bx(t)=(x_1(t),x_2(t),x_3(t))^\top:t\geq 0\}$, of the RRE in \eqref{eq:bifode1}.
Solving \eqref{eq:Code} and \eqref{eq:Vode} using the above matrices gives the matrices $\bC(s,t)$ and $\bD(s,t)$ required to derive the distribution of $\bxi(t)$ in $\eqref{eqn:xistate}$ and in turn the state $\bX(t)$ in $\eqref{eq:lnaansatzX}$.

\subsubsection{Accuracy of the LNA}\label{sec:lnaaccuracy}

The LNA accurately approximates the master equation in specific contexts. 
As discussed in SM (see Section~\ref{app:lna}),
the LNA is derived
using the Central Limit Theorem as $\Omega$ tends to $\infty$,  
providing an approximation of the probability distribution function satisfying the master equation. 
As a result, we expect the LNA to be accurate when $\Omega$ is sufficiently large.
The value of $\Omega$ needed to ensure accuracy depends on the characteristics of the system's dynamics and the length of time interval where the approximation is sought.  

For any given finite $\Omega$, if the LNA accurately represents the distribution of the state of a given system at some time-point, it is expected to maintain this accuracy for a subsequent short time interval (see \cite{Wallace2012}).
The duration for which the LNA remains accurate is influenced by the  dynamics of the system and the level of stochasticity present. 
For instance, it was observed in \cite{Minas2017} that for periodic systems with a moderate size $\Omega$, the time interval over which the LNA retains its accuracy is approximately equal to the period of the system.

\begin{figure}[h!]
    \centering
    \includegraphics[scale=1]{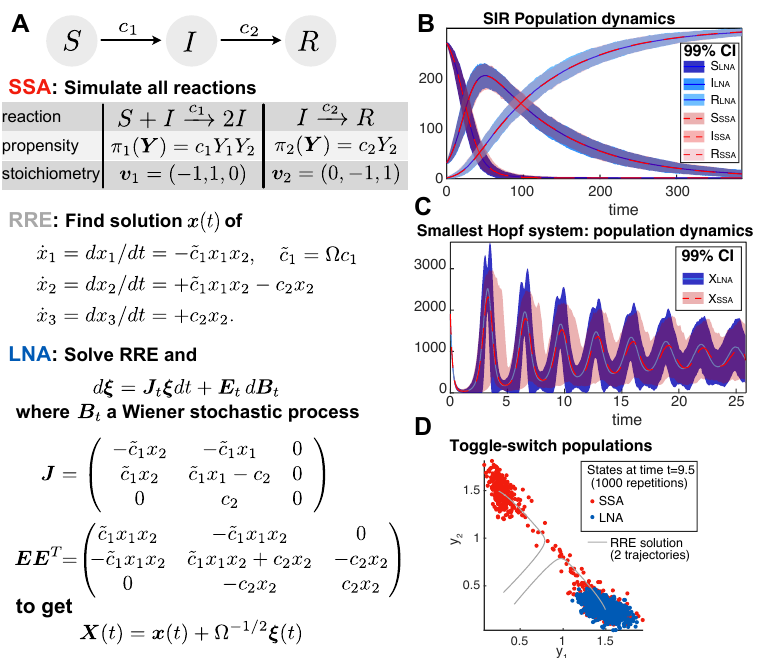}
    \caption{\label{fig:LNA} LNA method summary.
    A. LNA is a stochastic approximation of the stochastic population dynamics that can be exactly simulated using the SSA. LNA uses the numerical solutions of the deterministic RRE and a SDE that can be solved exactly. The SIR population dynamics model is used here as an example. 
    B. Comparison between SSA and LNA on the SIR population dynamics in terms of simulated $99\%$ Confidence Intervals (CI) shows astonishing agreement as the blue lines and shades almost completely overlap with the pink lines and shades. 
    C. In contrast, the same comparisons between the LNA and SSA in one of the three species of the oscillatory system in \cite{WILHELM1995} show significant discrepancies. 
    D. Significant discrepancies are also observed for the same comparisons in the bi-stable Genetic Toggle-Switch system in \cite{Gardner2000}.}
\end{figure}

In situations where the system displays linear dynamics (i.e.\ the propensity functions in \eqref{eq:prop} are linear with respect to $\bY(t)$) the first two moments of the LNA coincide with those of solutions of the master equation \cite{Lestas2008}. 
Additionally, empirical studies comparing the distribution of the system at specific time-points, as predicted by the LNA, with histograms of the state distribution derived by simulated trajectories using the SSA, have shown agreement when the initial conditions are in the vicinity of stable equilibrium points (see for instance, \cite{Thomas2013, Minas2017, Scott2006}, and Figure \ref{fig:LNA}(B)). 
Recent results identified the precise generic conditions under which the LNA is accurate near equilibrium points \cite{grunberg2024steinsmethodapproachlinear}.

It is important to note that, for moderate system sizes, the LNA tends to fail in approximating the long-term dynamics of multi-stable and oscillatory systems (see \cite{Scott2006, Boland2008, Ito2010, Thomas2013, Minas2017} and Figure \ref{fig:LNA}(C-D)).
We further discuss the failure of the LNA in Section~\ref{sec:phase}.

\section{The phase-corrected Linear Noise Approximation (pcLNA)}\label{sec:pcLNA}

In this section, we develop the phase-corrected Linear Noise Approximation (pcLNA) algorithm for stochastic simulation of population dynamics.
We begin by introducing a general definition of phase.

\subsection{Phase}\label{sec:phase}

Important in diagnosing the failure of the LNA approximation, and indeed the rest of this paper, is the idea of phase. 
Existing definitions of phase for stochastic systems typically extend the concept of phase for noisy limit cycles
\cite{SCHWABEDALJ.T.C2010Epdo,CaoAlexander2020APDE,EngelMaximilian2021ARDS}.
Here we introduce a definition suitable for any non-linear dynamics arising from the RRE \eqref{eq:bifode1}, including systems with oscillations or multi-stability. 

We use the term \emph{phase} in its common-language sense: a stage in a dynamic process.
To define the phase of an arbitrary point $\bX$ on a stochastic trajectory $\{\bX(t):t\geq 0\}$,
we compare it with a set of reference trajectories
\begin{equation}\label{eq:refTraj}
\Gamma=\bigcup_{\ell=1}^L\{\bx^{(\ell)}(t):t\ge 0\},\qquad L\ge 1,
\end{equation}
generated from the system in \eqref{eq:bifode1}.  
The choice of $\Gamma$ depends on the underlying dynamics and will be discussed in Section~\ref{sec:phaseCorrectionMap}.  
For example, if the deterministic system has an attracting limit cycle, an appropriate choice is a single trajectory lying on that cycle;  
if it has multiple stable equilibria, $\Gamma$ may consist of trajectories converging to each equilibrium.

We find an appropriate metric $d$ to measure the distance between $\bX$ and points on the trajectories in $\Gamma$. 
Using the metric $d$, we identify the point $\bG(\bX)\in\Gamma$ that is closest to $\bX$. 
The phase of $\bX$ is the time coordinate $s\geq 0$ of this closest point.  
Formally, the phase $s$ of $\bX$ is defined via
\begin{equation}\label{eq:phaseargmin}
    (j,s) = \argminB_{\ell\in \mathcal{L},\, t\geq 0} d(\bX,\bx^{(\ell)}(t)), \qquad \cL=\{1,\dots,L\},
\end{equation}
and the associated reference point is obtained through the map $\bG:\mathbb{R}^n\to\Gamma$,
\begin{equation}\label{eq:phasecorrect}
  \bG(\bX)=\bx^{(j)}(s).
\end{equation}

The most appropriate metric $d$ to be used for the phase definition above depends on the system dynamics and will be discussed in Section~\ref{sec:phaseCorrectionMap}. 
Aside from the phase-defining metric $d$ and the reference trajectories $\Gamma$ that are specific to the system dynamics, this definition of phase applies to any type of dynamics, not only oscillatory or multi-stable systems.

While the LNA is accurate for systems with linear dynamics (see Section \ref{sec:lnaaccuracy}), for systems with non-linear dynamics (e.g. oscillatory or multi-stable), as $t$ increases the phase of $\bX(t)$ will likely be very different to $t$. This means, with high probability, there exists some time point $\tau>0$ such that $\bX(\tau)$ is at a very different stage of its dynamic process to the deterministic $\bx(\tau)$ despite $\bX(0)=\bx(0)$. Since the drift and diffusion in equation \eqref{eq:langevin} which control the dynamics of $\bX(t)$ for $t\geq\tau$ under the LNA depend only on $\bx(t)$, the LNA will make predictions of $\bX(t)$ for $t\geq\tau$ as if it were in the same stage of its dynamic process as $\bx(t)$ rather than considering the location of $\bX(t)$. This leads to the LNA failing to capture the SSA dynamics of the reaction network $\{\bY(t)/\Omega :   t\geq 0\}$ where $\{\bY(t) :   t\geq 0\}$ evolves according to \eqref{eq:rtc}. We call this phenomenon a phase drift.

Therefore, initialising the LNA to the correct phase is a necessary condition for the LNA to be accurate for non-linear systems. 
This leads us to speculate that, if one can continually initialise the LNA with the correct phase so that the phase of the stochastic and deterministic trajectories remain in agreement, then the LNA will remain accurate at all times and produce paths agreeing with those from the SSA. 
We argue that, for some classes of dynamical systems, this is indeed the case.
For instance, it is shown in \cite{Minas2017} that this is true for systems whose RRE has an attractive limit cycle solution and a range of moderate system sizes.

This is the basis of the simulation algorithm developed next. 
We adopt the term used in \cite{Minas2017} to name this simulation algorithm, phase-corrected Linear Noise Approximation (pcLNA).

\subsection{The pcLNA algorithm}\label{pcLNAsect}

Here we present the phase-corrected Linear Noise Approximation (pcLNA) algorithm for simulating reaction networks with non-linear dynamics using LNA transitions.
For each $t\geq 0$, we define the pcLNA ansatz;
\begin{equation}\label{eq:pcLNA}
    \bX(t) = \bx(t) + \Omega^{-1/2} \bxi(t) = \bG(\bX(t)) + \Omega^{-1/2}\bkappa(t).    
\end{equation}

Here $\bG:\bbR^n\to \Gamma$, is the phase-correcting 
function \eqref{eq:phasecorrect} that maps any stochastic state $\bX(t)\in \bbR^n$ to a point lying on a trajectory in some set of reference RRE solutions $\Gamma=\bigcup_{\ell=1}^L\{\bx^{(\ell)}(t):t\geq 0\}$ with its corresponding time being the phase $s\geq 0$ of the stochastic state $\bX(t)$. 
The term $\bkappa(t)$ denotes the deviation of $\bX(t)$ away from $\bG(\bX(t))$ scaled by $\Omega^{-1/2}$. We use this ansatz to simulate stochastic trajectories as described in Algorithm \ref{alg:gen_pclna}.

\begin{algorithm}
\caption{General pcLNA algorithm}\label{alg:gen_pclna}
\textbf{Inputs:}\\
a.\ Initial conditions: $t_0=0$, $\bX(0)$, $\bx^{(\ell)}(0)$, $\ell=1,\dots,L$. Parameters: $\Delta t$, $\Omega$.\\ 
b.\ Solutions $\Gamma = \bigcup_{\ell=1}^L\{\bx^{(\ell)}(t) :   t\geq 0\}$ of a RRE as in \eqref{eq:bifode1} and the corresponding solutions $\bC^{(\ell)}$ and $\bD^{(\ell)}$ of \eqref{eq:Code}, \eqref{eq:Vode}, respectively.\\
c.\ A metric $d$ on $\bbR^n$ for the phase-correcting map $\bG:\bbR^n\to \Gamma$ as in \eqref{eq:phaseargmin}, \eqref{eq:phasecorrect}.\\[0.5em]
\textbf{Steps:}
Set $(j_0,s_0)$ such that $\bG(\bX(0))=\bx^{(j_0)}(s_0)$ and take $\bkappa(s_0)=\Omega^{1/2}(\bX(0)-\bx^{(j_0)}(s_0))$. Then, for $i=1,2,\dots$,
\begin{enumerate}
    \item draw sample $\bxi(s_{i-1}+\Delta t)$ from the MVN distribution with mean $\bC^{(j_{i-1})}(s_{i-1},s_{i-1}+\Delta t)\bkappa(s_{i-1})$, and covariance matrix $\bD^{(j_{i-1})}(s_{i-1},s_{i-1}+\Delta t)$; \label{alg:gen_pclna_step1} 
    \item compute $\bX(t_i) = \bx^{(j_{i-1})}(s_{i-1}+\Delta t) + \Omega^{-1/2}\bxi(s_{i-1}+\Delta t)$;  \label{alg:gen_pclna_step2}
    \item set $(j_i,s_i)$ such that $\bG(\bX(t_i))=\bx^{(j_i)}(s_i)$;
    \label{alg:gen_pclna_step3}
    \item set $\bkappa(s_i) = \Omega^{1/2}(\bX(t_i)-\bx^{(j_i)}(s_i))$.  \label{alg:gen_pclna_step4}
\end{enumerate}
\textbf{Outputs:} $\{\bX(t_i) :   i=0,1,2,\dots \}$.
\end{algorithm}

\begin{figure}[h!]
    \centering
    \includegraphics[scale=1]{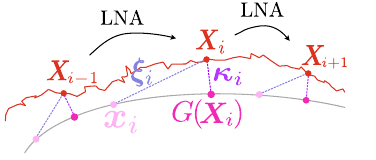}
    \caption{\label{fig:gpcLNA} General pcLNA algorithm over steps $i-1,i,i+1$. For simplicity we illustrate the case where $j_{i-1}=j_i=j_{i+1}$, so the same RRE solution is used over these steps.
    pcLNA proceeds with standard LNA steps interrupted by phase corrections with $\bx^{(j_i)}(t_i)$ replaced by $\bG(\bX(t_i))=\bx^{(j_i)}(s_i)$, and the perturbations $\bxi(t_i)$ replaced by  $\bkappa(s_i)$. 
    }
\end{figure}

The Algorithm \ref{alg:gen_pclna} uses the standard LNA transition in steps \ref{alg:gen_pclna_step1}-\ref{alg:gen_pclna_step2} interrupted by phase-correcting maps in steps \ref{alg:gen_pclna_step3}-\ref{alg:gen_pclna_step4}. 
The map $\bG$ is used to find the phase for $\bX(t_i)$. 
After phase correction, the phase of the system is moved from $s_{i-1}+\Delta t$ to $s_i$, the deterministic state moved from $\bx^{(j_{i-1})}(s_{i-1}+\Delta t)$ to $\bx^{(j_i)}(s_i)$ and the fluctuations $\bxi(s_{i-1}+\Delta t)$ are replaced with the remaining fluctuations 
after phase correction, $\bkappa(s_i)$.
Provided $\Delta t$ is not too large, phase drifts occurring in the next LNA transition will be small. 
Indeed, $\{\bkappa(s_i) :   i\in\bbNz\}$ should have uniformly bounded variance and remain sufficiently small for the LNA predictions to be valid over the short time intervals $[s_i,s_i+\Delta t), \ i\in\bbNz$.
For computational efficiency, it is important that the same solutions of ODEs $\eqref{eq:bifode1}$, $\eqref{eq:Code}$ and $\eqref{eq:Vode}$ are used in all steps. Because these ODEs only need to be solved once before any simulations, the computational speed is significantly increased. The steps are illustrated in Figure \ref{fig:gpcLNA}.

All that is preventing one from readily applying the steps in the pcLNA algorithm is an explicit phase-defining metric $d$ and a set of reference trajectories $\Gamma$ used in the phase-correcting map $\bG$ as in \eqref{eq:phaseargmin}, \eqref{eq:phasecorrect}. 
A natural question arises: 
does there exist a single choice of metric $d$ and reference set $\Gamma$ that can be used for all reaction networks to produce accurate simulations over long times?
As we explain below, the answer is no. Phase depends on the system’s underlying dynamics, and in particular on which components generate persistent non-linear behaviour. Consequently, no universal metric can correctly capture phase across qualitatively different systems.

In the next section, we show how tools from dynamical systems theory provide a way to determine a phase-defining metric $d$ and the set of reference trajectories $\Gamma$ tailored to the system’s qualitative dynamics.

\subsection{Choosing the phase-defining metric $d$ and set of reference trajectories $\Gamma$} \label{sec:phaseCorrectionMap}

We use three connected concepts from dynamical systems theory to guide our approach in choosing a metric $d$ and set $\Gamma$ used in the phase-correcting map $\bG$.
The first concept is topological equivalence, which allows us to group systems exhibiting qualitatively identical dynamics. 
The second concept is decomposition, enabling us to break down complex (potentially high-dimensional) systems into a (possibly low-dimensional) non-linear component with persistent dynamics and components with transient dynamics. 
The third concept involves exemplar systems within each group that are well understood and can serve as references. 
We highlight the aspects of these concepts that are used in our approach in Sections~\ref{sec:topequi}, \ref{sec:decomp} and \ref{sec:normform} and explain how we practically apply them in the construction of a suitable metric $d$ and set $\Gamma$ in Section \ref{sec:phaseCorrConseq}.
We will then use this approach in Sections \ref{sec:hopf} and \ref{sec:bi-stable} to define $d$ and $\Gamma$ for systems presenting a Hopf bifurcation 
and different choices of $d$ and $\Gamma$ for systems exhibiting bi-stability through a fold bifurcation. 

\subsubsection{Topological equivalence}\label{sec:topequi}

Two continuous dynamical systems of the form \eqref{eq:bifode1} are topologically equivalent if their dynamics are qualitatively identical. Formally, this holds if there exists a homeomorphism $\bh$ which maps paths of one system onto paths of the other while preserving the direction of time.
Local topological equivalence (l.t.e.) occurs when this holds in neighborhoods of the systems’ domains (see Definitions 2.14-5 in \cite{Kuznetsov2011}, or SM Section~\ref{SIsec:man}).  

A practical approach to identify l.t.e.\ is to examine the eigenvalues of the
Jacobian in \eqref{eq:jac} near an equilibrium point $(\bx_{\alpha_p},\alpha_p)$, with $\bF(\bx_{\alpha_p},\alpha_p)=\bzero$. 
Let $n_0$, $n_+$, and $n_-$ be the number of eigenvalues of $\bJ(\bx_{\alpha_p},\alpha_p)$ that have zero, positive, and negative real part, respectively.
An equilibrium is hyperbolic if $n_0=0$; otherwise, it is non-hyperbolic. For hyperbolic equilibria, l.t.e.\ is ensured if $n_+$ and $n_-$ match between systems (see Theorem 2.2 in \cite{Kuznetsov2011}); the non-hyperbolic case is more subtle and discussed below.

\subsubsection{Decomposition}\label{sec:decomp}

Using results from centre manifold theory (see \cite{Kuznetsov2011,ShoshitaishviliA.N.1972Bott,Takens1971,Takens1983}) and It\^{o}'s differential calculus (see \cite{LeGallsdebook}), we derive a novel decomposition of the stochastic process generated by the LNA into components with distinct dynamical behaviour.
For a detailed description of the results used in this section and the calculations performed, see SM Section~\ref{SIsec:man}.

Suppose the system \eqref{eq:bifode1} exhibits an equilibrium $(\bx_\alpha,\alpha)$ which is non-hyperbolic ($n_0>0$) when $\alpha=\alpha_p$. In this setting, the system's non-linear dynamics can be reduced to occur on an $\alpha$-dependent $n_0$-dimensional \emph{centre manifold} tangent to the eigenspace corresponding to the eigenvalues of $\bJ(\bx_{\alpha_p},\alpha_p)$ with zero real part (see Theorem 5.1 and Lemma 5.1 in \cite{Kuznetsov2011}).

In fact, the system is l.t.e.\ to
    \begin{equation}\label{eq:Takensform}
    \left\{
    \begin{aligned}
        \dot{\bu} &= \bm f(\bu,\alpha), \\
        \dot{\bv} &= \bA_{\text{s}}(\bu,\alpha)\bv,\\
        \dot{\bw} &=\bA_{\text{u}}(\bu,\alpha)\bw,
    \end{aligned}
    \right.
\end{equation}
for paths and parameter values near to $(\bx_{\alpha_p},\alpha_p)$ where for each $t\geq 0$,
$\bu(t)\in \bbR^{n_0}$, $\bv(t)\in \bbR^{n_-}$, $\bw(t)\in \bbR^{n_+}$, $\bm f:\bbR^{n_0}\times\bbR\to\bbR^{n_0}$ is smooth,
$\bA_{\text{s}}(\bu(t),\alpha)$ is an $n_-\times n_-$ (stable) matrix with all of its eigenvalues having negative real part,
and $\bA_{\text{u}}(\bu(t),\alpha)$ is an $n_+\times n_+$ (unstable) matrix with all of its eigenvalues having positive real part (see Section 4 in \cite{Takens1971} or Section 2 in \cite{Takens1983}).
Moreover, if the respective local homeomorphism denoted by $\bh_\alpha:\bbR^n\to \bbR^n$ satisfies a set of generic nonresonance conditions 
of the eigenvalues of $\bJ(\bx_{\alpha_p},\alpha_p)$ that hold almost surely (see SM Section~\ref{SIsec:man}), $\bh_\alpha$ is at least twice differentiable in a neighborhood of $\bx_{\alpha_p}$.
In this setting, It\^{o}'s lemma (see \cite[Theorem 5.10]{LeGallsdebook}) can be applied locally to characterise the image of the LNA under $\bh_\alpha$:
\begin{equation}
    \begin{pmatrix}
        \bU(t) \\ \bV(t) \\ \bW(t)
    \end{pmatrix}
    := \bh_\alpha(\bx(t) + \Omega^{-1/2}\bxi(t)) =
     \begin{pmatrix}
        \bu(t) \\ \bv(t) \\ \bw(t) 
    \end{pmatrix}
    + \Omega^{-1/2}
    \begin{pmatrix}
        \bm \xi_{\bu}(t) \\ \bm \xi_{\bv}(t) \\ \bm \xi_{\bw}(t)
    \end{pmatrix}
    + O(\Omega^{-1}),
\end{equation}
for times $t\geq 0$ where $\bx(t) + \Omega^{-1/2}\bxi(t)$ remains close to $\bx_{\alpha_p}$. Here $(\bu(t),\bv(t),\bw(t))^\top := \bh_\alpha(\bx(t))$ evolves according to \eqref{eq:Takensform} and
the perturbation $\tilde{\bm \xi}(t) = (\bm \xi_{\bu}(t),\bm \xi_{\bv}(t),\bm \xi_{\bw}(t))^\top$ evolves according to the system of SDEs
\begin{equation}\label{eq:eigenSDE}
    \begin{aligned}
    d \tilde{\bm \xi} = \tilde{\bm J}_t \tilde{\bm \xi} dt + \tilde{\bm E}_t dB_t 
    \end{aligned}
\end{equation}
where
\begin{equation}
\tilde{\bJ}_t =
    \begin{bmatrix}
        \nabla_\bu \bm f(\bu,\alpha) & 0 & 0\\
        \nabla_\bu (\bA_{\text{s}}(\bu,\alpha)\bv) & \bA_{\text{s}}(\bu,\alpha) & 0 \\
        \nabla_\bu (\bA_{\text{u}}(\bu,\alpha)\bw) & 0 & \bA_{\text{u}}(\bu,\alpha) 
    \end{bmatrix}
    ,\quad \tilde{\bm E}_t = \nabla_\bx \bh_\alpha(\bx(t))\bE_t 
\end{equation}

Under this local homeomorphism, the LNA, omitting $O(\Omega^{-1})$ terms, can be decomposed into 3 processes: 
\begin{enumerate}
\item $\bU(t)\approx\bu(t) + \Omega^{-1/2}\bm \xi_\bu(t)$ the sum of a non-linear deterministic process and a stochastic process with drift  $\nabla_\bu \bm f(\bu,\alpha)\bm \xi_u$;
\item $\bV(t) \approx \bv(t) + \Omega^{-1/2}\bm \xi_\bv(t)$ the sum of an exponentially decaying deterministic process and a linear stochastic process with stable drift matrix, 
\item $\bW(t)\approx\bw(t) + \Omega^{-1/2}\bm \xi_{\bw}(t)$ the sum of an exponentially growing deterministic process and a linear stochastic process with unstable drift matrix. 
\end{enumerate}
That is, at the $O(\Omega^{-1/2})$ scale, 
$\bV(t)$ quickly drifts toward zero and $\bW(t)$ quickly drifts outside the neighborhood, leaving all of the effective stochastic behaviour of the LNA to be described by $\bU(t)$. This implies that any phase drift responsible for the long-term failure of the LNA arises from the dynamics of $\bU(t)$ relative to $\bu(t)$; the other processes, $\bV(t)$ and $\bW(t)$, play no long-term role and do not contribute to persistent phase drift.

Throughout the rest of the paper, we refer to the first $n_0$ coordinates of the image of $\bh_\alpha$, where $\bU(t)$ and $\bu(t)$ evolve, as the centre coordinates; the next $n_-$ coordinates, where $\bV(t)$ and $\bv(t)$ evolve, as the stable coordinates; and the last $n_+$ coordinates, where $\bW(t)$ and $\bw(t)$ evolve, as the unstable coordinates.

While this mapping is generally non-linear, its action on the \emph{centre coordinates} admits a first-order approximation near the equilibrium (See SM Section~\ref{SIsec:man}). Specifically, for any point $\bX$ sufficiently close to $\bx_{\alpha_p}$, its location $\bU$ in the centre coordinates of $\bh_\alpha(\bX)$ satisfies
\begin{equation}
\bU = \bm P_c(\bX-\bx_{\alpha_p}) + O(\norm{\bX-\bx_{\alpha_p}}^2),
\end{equation}
where $\bm P_c$ denotes the coordinate projection into the basis of the eigenspace of $\bJ(\bx_{\alpha_p},\alpha_p)$ corresponding to eigenvalues of zero real part. Explicitly, $\bm P_c:=({\bR_c}^\top\bR_c)^{-1}{\bR_c}^\top$ and $\bR_c$ is the $n\times n_0$ matrix whose columns form an eigenbasis corresponding to the eigenvalues of $\bJ(\bx_{\alpha_p},\alpha_p)$ with zero real part. It is important to note this is a first-order approximation of these coordinates and valid only sufficiently close to the equilibrium.
Consequently, given a realization of the LNA, $\bX(t)=\bx(t)+\Omega^{-1/2}\bxi(t)$, the locations of $\bU(t)$ are approximately
\begin{equation}
    \bU(t) \approx \bm P_c (\bX(t)-\bx_{\alpha_p}).
\end{equation}
Similarly,
\begin{equation}
    \bu(t) \approx \bm P_c (\bx(t)-\bx_{\alpha_p}).
\end{equation}
  
\subsubsection{Normal forms}\label{sec:normform}

\begin{figure}[h!]
    \centering
    \includegraphics[scale=1]{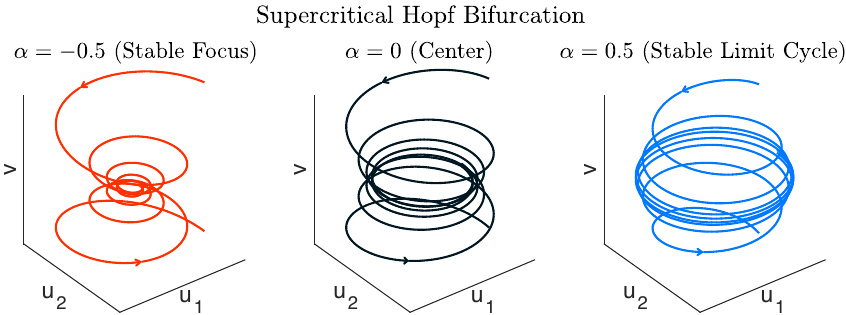}
    \caption{\label{fig:hopf} The dynamics of a generic normal form of the supercritical Hopf bifurcation. The two-dimensional centre manifold has coordinates $\bu=(u_1,u_2)$, while the third coordinate $v$ satisfies $\dot{v}=-v$. Selected orbits for the three different cases are shown. 
    }
\end{figure}

Normal forms are simple exemplar systems representing classes of l.t.e.\ systems.  
Low-dimensional centre manifolds have well-studied normal forms, including the fold and Hopf bifurcations (see \cite{Kuznetsov2011}, Figure \ref{fig:hopf}). We use normal
forms in two ways. 
First, to specify the conditions under which a given
system is l.t.e.\ to the normal form, typically related to the eigen-decomposition of the Jacobian.
Secondly, we use their simpler form to identify the key qualitative features (e.g., oscillations, multi-stability) shared by systems in their class, guiding the choice of an appropriate metric $d$ and set $\Gamma$ for the phase-correcting map $\bG$.

\subsubsection{Procedure for constructing the phase-correcting map} \label{sec:phaseCorrConseq}

The theory in Sections \ref{sec:topequi}-\ref{sec:normform} leads us to the following principles for defining the phase-defining metric $d$ and set of reference trajectories $\Gamma$:
\begin{enumerate}[label=\Alph*.]
    \item Local topological equivalence (l.t.e.) identifies classes of systems whose trajectories are qualitatively identical near equilibria allowing the same construction of the phase-defining metric $d$ and reference set $\Gamma$ to apply across all systems in the class.
    \item Decomposition isolates the coordinates in which non-linear, persistent dynamics — and hence phase drifts — occur.
    Phase and phase correction are defined exclusively in these centre coordinates, while the transient dynamics of the stable and unstable coordinates are disregarded when constructing the metric $d$.
    \item Normal forms provide canonical low-dimensional representatives of each l.t.e. class and make explicit the geometric structure of trajectories in the centre coordinates. This structure determines both the appropriate phase-defining metric and the choice of representative reference trajectories $\Gamma$.
\end{enumerate}

\begin{figure}[h!]
    \centering
    \includegraphics[width=\textwidth]{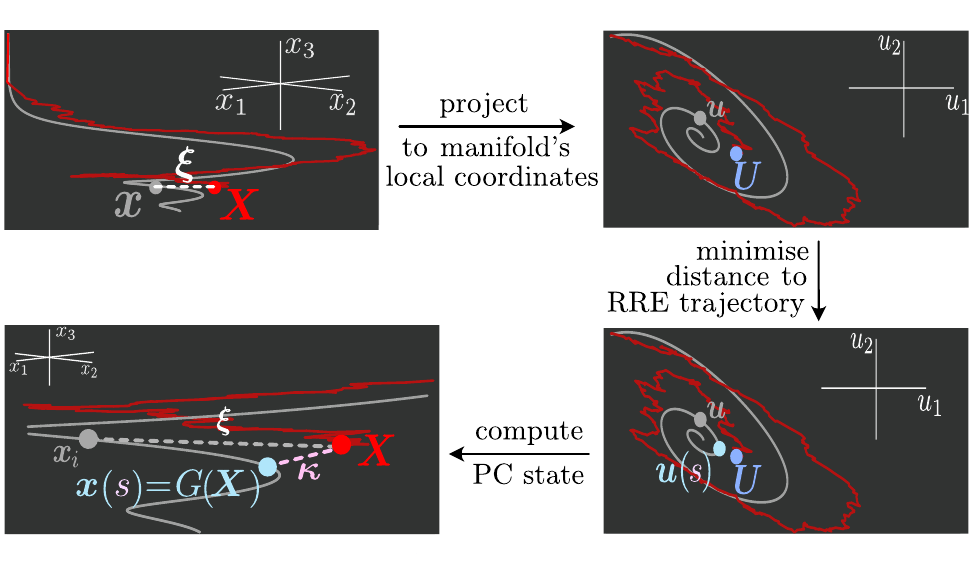}
    \caption{\label{fig:Gmap} Phase correction.
    The phase-correcting map $\bG(\bX_i)$ for a point $\bX_i$ on the stochastic trajectory is computed in three steps. First, the RRE solution and point $\bX_i$ is projected into the centre coordinates of the non-linear centre manifold. 
    Then, the point, $\bu(s_i)$ lying on the projected RRE trajectory $\{\bu(t):t\geq 0\}$ that minimises the distance to $\bU_i$ is found. Finally, the corresponding point, $\bx(s_i)$, with the same time-point $s_i$ with $\bu(s_i)$ and that lies on the RRE solution $\{\bx(t):t\geq 0\}$ is found. The point $\bx(s_i)$ and the perturbation $\bkappa_i$ (instead of $\bx_i$ and $\bxi_i$) are used in the next LNA transition. The magnitude of $\bkappa_i$ is smaller than or equal to the magnitude of $\bxi_i$ on the centre coordinates. 
    }
\end{figure}

Using these principles we define a procedure for constructing the phase-correcting map $\bG$ for any system as in \eqref{eq:bifode1}.
\begin{enumerate}
    \item Check if there exists a parameter value $\alpha_p$ and a state $\bx_{\alpha_p}$ in a neighborhood of interest 
    such that $\bF(\bx_{\alpha_p},\alpha_p)=\bzero$, and the      
    Jacobian $\bJ(\bx_{\alpha_p},\alpha_p)$ has $n_0>0$ eigenvalues with zero real part. If so, proceed to the next step; otherwise, consider using an alternative method (including the standard LNA).
    \item Identify the local topological equivalence (l.t.e.) class of \eqref{eq:bifode1} by
    checking whether it satisfies the conditions of a known normal form,
    such as those associated with Hopf bifurcations or bi-stable systems 
    (see Sections~\ref{sec:hopf} and~\ref{sec:bi-stable}). 
    \item 
    Based on the l.t.e.\ class dynamics and, where applicable, the corresponding normal form, choose a set of reference trajectories $\Gamma=\bigcup_{\ell=1}^L\{\bx^{(\ell)}(t): t\ge0\}$, containing at least one deterministic solution of each distinct type of qualitative behaviour.
    For example, if the system admits three stable equilibria, select three
    trajectories converging to each equilibrium.
    \item Define a projection map to find the approximate location of a point near $\bx_{\alpha_p}$ in the centre coordinates:
    \begin{equation}\label{eq:projGen}
      \text{proj}: \bbR^n \to \bbR^{n_0}, \quad \text{proj}(\bx) = \bm P_c (\bx-\bx_{\alpha_p}),
     \end{equation} 
    where $\bm P_c=({\bR_c}^\top\bR_c)^{-1}{\bR_c}^\top$ and $\bR_c$ is an $n\times n_0$  
    matrix whose columns form an eigenbasis corresponding to the eigenvalues of $\bJ(\bx_{\alpha_p},\alpha_p)$ that have zero real part. 
    \item 
    Compute the projections of all reference trajectories
    \begin{equation}\label{eq:projTraj}
    \{\bu^{(\ell)}(t)=\text{proj}(\bx^{(\ell)}(t)) :   t\geq 0\}, \quad \ell\in \cL, \,\,\cL:=\{1,\dots,L\}    
    \end{equation}
    and, for any stochastic state $\bX\in \bbR^n$, define
    \begin{equation}\label{eq:proj}
    \bU:=\text{proj}(\bX).
    \end{equation}
    \item Define the distance metric $d:\bbR^n\times\bbR^n\to[0,\infty)$, by
    \begin{equation}\label{eq:dmetric}
        d(\by,\bx) := \|\text{proj}(\by)-\text{proj}(\bx)\|.
    \end{equation}
    \item 
    Using \eqref{eq:projGen}-\eqref{eq:dmetric}, the phase-correcting map $\bG:\bbR^n \to \Gamma$ defined in \eqref{eq:phaseargmin} is given by
    \begin{equation}\label{eq:phasemap}  \bG(\bX)=\bx^{(j)}(s),\,\text{ where }
    (j,s)=\argminB_{\ell\in\cL,\,t\geq 0} d(\bX,\bx^{(\ell)}(t)) = \argminB_{\ell\in\cL,\,t\geq 0} \| \bU-\bu^{(\ell)}(t)\|.
    \end{equation}
\end{enumerate}
In summary, this procedure defines the phase-correcting function $\bG$, which maps any stochastic state $\bX$ to the point on the reference trajectories in $\Gamma$ that is closest to $\bX$ in the approximate centre coordinates.
Figure \ref{fig:Gmap} illustrates steps (5-7) above. 

Although the true centre coordinates are given by the $\alpha$-dependent map $\bh_\alpha$, the linear projection \eqref{eq:projGen} provides a first-order approximation that is independent of $\alpha$ and can therefore be
used consistently across parameter values.

For the numerical investigations in Section~\ref{sec:numerical_investig}, we use
the Euclidean norm \eqref{eq:dmetric}, though other metrics may be employed.

In Sections \ref{sec:hopf} and \ref{sec:bi-stable} we apply this procedure for two concrete types of dynamics: Oscillations (sustained or damped) and bi-stability, which occur when the RRE in \eqref{eq:bifode1} contains a Hopf bifurcation and generic fold bifurcation, respectively. 
We show the resulting metrics $d$ and sets $\Gamma$ can be used to accurately simulate such dynamics in Section \ref{sec:numerical_investig}.

\section{Hopf bifurcation systems}\label{sec:hopf}

We now consider systems which present oscillatory behaviour. 
In particular, we aim to generate a stochastic trajectory $\{\bX(t_i):i=0,1,2,\dots\}$, with $\bX(t_i)\in \bbR^n$ for a reaction network with macroscopic RRE as in \eqref{eq:bifode1} and Jacobian $\bJ(\bx,\alpha)$ as in \eqref{eq:jac} 
    that satisfy the following conditions:
    \begin{enumerate}[label=(H.\Roman*)]
    \item there exists an $\alpha_p\in\bbR$ and $\bx_{\alpha_p}\in\bbR^n$ such that,
    $ \bF(\bx_{\alpha_p},\alpha_p) = \bzero, $ \label{HI}
    \item for $\alpha$ local to $\alpha_p$, all the eigenvalues of the Jacobian matrix, $\bJ(\bx_{\alpha},\alpha)$ 
    are negative except for a pair of complex conjugate eigenvalues $\lambda(\alpha), \ \bar{\lambda}(\alpha)$ such that
     \[ \lambda(\alpha) = \mu(\alpha) + i \omega(\alpha), \quad \text{where} \quad \mu(\alpha_p) = 0, \ \omega(\alpha_p) > 0.\] \label{HII}
\end{enumerate}
Condition \ref{HI} is precisely the requirement for the macroscopic reaction rate equation to exhibit an equilibrium. 
Condition \ref{HII} ensures the equilibrium is non-hyperbolic and therefore prescribes the existence of a parameter dependent, two-dimensional, attracting manifold near $(\bx_{\alpha_p},\alpha_p)$ for which \eqref{eq:bifode1} is l.t.e. to a system
\begin{equation}
        \left\{
    \begin{aligned}
        \dot{\bu} &= \bm f(\bu,\alpha), \\
        \dot{\bv} &= \bA(\bu,\alpha)\bv,\\
    \end{aligned}
    \right.
\end{equation}
where for each $t\ge 0$, $\bu(t)\in\bbR^2, \ \bv(t)\in\bbR^{n-2}, \ \bm f:\bbR^2\times\bbR\to\bbR^2$ is smooth and $\bA(\bu(t),\alpha)$ is an $(n-2)\times(n-2)$ matrix with all of its eigenvalues having negative real part. See Figure~\ref{fig:hopf} for an illustration of the paths of the normal form of this system.

Since all the eigenvalues other than the complex pair have negative real parts, the calculations in Section \ref{sec:decomp} reveal the image of the LNA, $\bX(t) = \bx(t) + \Omega^{-1/2}\bxi(t)$, under the local homeomorphism gives two stochastic processes: $\bU(t)\approx \bu(t) + \Omega^{-1/2}\bm \xi_\bu(t)$, evolving in the centre coordinates and exhibiting the non-linear dynamics, and $\bV(t)\approx \bv(t)+\Omega^{-1/2}\bm \xi_\bv(t)$, evolving in the stable coordinates and decaying at an exponential rate. Their stochastic perturbations, $\bm \xi_\bu(t)$ and $\bm \xi_\bv(t)$, evolve according to \eqref{eq:eigenSDE}.

\subsection{The pcLNA algorithm for Hopf bifurcation systems}\label{sec:phaseHopf}

For a reaction network with macroscopic RRE as in \eqref{eq:bifode1} satisfying conditions \ref{HI}, \ref{HII}, we now aim to define the projection metric $d$ and set of reference trajectories $\Gamma$ to define the phase-correcting map $\bG$, as described in Section \ref{sec:phaseCorrConseq}.
Consider some value $\alpha$ of the bifurcation parameter that is near the critical $\alpha_p$. 
The normal form dynamics imply that trajectories starting from different initial conditions rapidly converge to a common path. 
This motivates the use of a single deterministic trajectory $\Gamma=\{\bx(t) :   t \geq 0 \}$ near $\bx_{\alpha_p}$, which proved sufficient to achieve accurate simulations for all systems considered in Section \ref{sec:numerical_investig}.

To obtain a readily applicable $\bG$ using the steps in Section \ref{sec:phaseCorrConseq} we identify the conjugate pair of eigenvectors, $\hat{\bu}_1\pm i \,\hat{\bu}_2$, corresponding to the imaginary eigenvalues of the Jacobian $\bJ(\bx_{\alpha_p},\alpha_p)$ as in \ref{HII}. 
Then, the eigenspace corresponding to these complex eigenvalues is spanned by $\{\hat{\bu}_1,\hat{\bu}_2\}$.
Letting $\bR_c = [\hat{\bu}_1 \; \hat{\bu}_2]$ we define the map $\pr: \bbR^n \to \bbR^2$, $\pr(\bx)=\bm P_c(\bx-\bx_{\alpha_p})$ where $\bm P_c=({\bR_c}^\top\bR_c)^{-1}{\bR_c}^\top$. 
Now $d$ and $\bG$ can be defined as in \eqref{eq:dmetric}, and \eqref{eq:phasemap} with $L=1$; phase correction occurs to minimize the distance in the approximate centre coordinates between the stochastic state $\bU=\pr(\bX)$ and single deterministic path $\{\bu(t)=\pr(\bx(t)):t\geq 0\}$. Having defined the metric $d$ for the map $\bG$, one may follow the steps in Algorithm \ref{alg:gen_pclna} to produce simulations of Hopf bifurcation systems.

\begin{figure}[h]
    \centering
    \includegraphics[width=\textwidth]{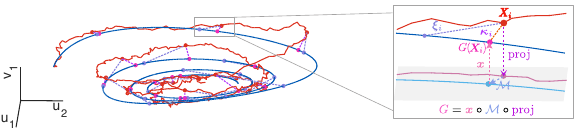}
    \caption{\label{fig:HopfPC} Phase correction (PC) for Hopf bifurcation systems. (Left) The stochastic trajectory (red) and deterministic trajectory (blue) in coordinates $(\bu,v_1)$ are displayed. 
    The stochastic state (red circle) and the perturbations before ($\bxi$, purple, dashed line) and after ($\bkappa$, pink, dashed line) PC are displayed. (Right) Zoom in of a PC with the $\bG$ map analysed as a composition of the projection ($\pr$), distance minimisation ($\mathcal{M}$), and lift back to original coordinates ($\bx$) functions.
    }
\end{figure}

Since only one RRE solution $\Gamma$ is required for this type of system the map $\bG$ can be written as the composition $\bG = \bx \circ \mathcal{M} \circ \pr$ where $\mathcal{M}:\bbR^2 \to [0,\infty)$, such that $s=\mathcal{M}(\bU)= \argminB_{t\geq 0}\norm{\bU-\bu(t)}$ (see Figure \ref{fig:HopfPC}). For ease, we provide a pcLNA algorithm specifically for Hopf bifurcation systems using this notation in Algorithm \ref{alg:Hopf_pclna}. In Section \ref{sec:numerical_investig}, we show that this algorithm 
is sufficient to 
achieve high levels of accuracy.

\begin{algorithm}[h]
\caption{pcLNA for Hopf bifurcation systems}\label{alg:Hopf_pclna}
\textbf{Inputs:}\\
a.\ Initial conditions: $t_0=0$, $\bx(0)$, $\bX(0)$, Parameters: $\Delta t$, $\Omega$.\\ 
b.\ System satisfying \ref{HI}, \ref{HII}, with solution $\Gamma=\{\bx(t) :   t\geq 0\}$ of \eqref{eq:bifode1} and the corresponding solutions $\bC$ and $\bD$ of \eqref{eq:Code}, \eqref{eq:Vode}, respectively.\\
c. The projection map $\pr:\bbR^n\to \bbR^2$, $\pr(\bx)=\bm P_c(\bx - \bx_{\alpha_p})$ where $\bm P_c=({\bR_c}^\top\bR_c)^{-1}{\bR_c}^\top$ and $\bR_c = [\hat{\bu}_1 \; \hat{\bu}_2]$ with $\hat{\bu}_1$, $\hat{\bu}_2$ an eigenbasis corresponding to the conjugate pair of imaginary eigenvalues of $\bJ(\bx_{\alpha_p},\alpha_p)$.\\
\textbf{Steps:}
Compute $\{\bu(t):t\geq 0\}$, where $\bu(t)=\pr(\bx(t))$ and set $s_0$ such that $\bG(\bX(0))=\bx(s_0)$ with $\bkappa(s_0)=\Omega^{1/2}(\bX(0)-\bx(s_0))$.\\  
For $i=1,2,\dots$,
\begin{enumerate}
    \item draw sample $\bxi(s_{i-1}+\Delta t)$ from the MVN distribution with mean $\bC(s_{i-1},s_{i-1}+\Delta t)\bkappa(s_{i-1})$, and covariance matrix $\bD(s_{i-1},s_{i-1}+\Delta t)$;
    \item compute $\bX(t_i) = \bx(s_{i-1}+\Delta t) + \Omega^{-1/2}\bxi(s_{i-1}+\Delta t)$;
    \item compute $\bU_i=\pr(\bX(t_i))$, and find $s_i = \mathcal{M}(\bU_i) = \argminB_{t\geq 0}\norm{\bU_i-\bu(t)}$.
    \item compute $\bkappa(s_i) = \Omega^{1/2}(\bX(t_i)-\bx(s_i))$.
\end{enumerate}
\textbf{Outputs:} $\{\bX(t_i) :   i=0,1,2,\dots \}$.
\end{algorithm}

\section{Bi-stable systems}\label{sec:bi-stable}

We now consider systems that present bi-stability and aim to derive stochastic trajectories $\{\bX(t_i) : i= 0,1,2,\dots\}$, with $\bX(t_i)\in \bbR^n$, for a reaction network whose RRE limit (as $\Omega\to \infty$) satisfying \eqref{eq:bifode1} and Jacobian, $\bJ(\bx,\alpha)$, as in \eqref{eq:jac} that satisfy the following conditions:
\begin{enumerate}[label=(B.\Roman*)]
    \item there exists an $\alpha_p\in \bbR$ and $\bx_{\alpha_p} \in \bbR^n$ such that $\bF(\bx_{\alpha_p},\alpha_p)=\bzero$;\label{BI}
    \item all the eigenvalues of the Jacobian matrix $\bJ(\bx_{\alpha_p},\alpha_p)$ are negative except for one zero eigenvalue.\label{BII}
    \item for $\alpha$ near $\alpha_p\in \bbR$ there exists $\bx^{(1)}_* \in \bbR^n$ such that $\bF(\bx^{(1)}_*,\alpha)=\bzero$ and all the eigenvalues of the Jacobian matrix $\bJ(\bx^{(1)}_*,\alpha)$ are negative. \label{BIII}
\end{enumerate}

The last condition (B.III) ensures the presence of a stable equilibrium point for all values of $\alpha$ near $\alpha_p$.
The conditions (B.I-II) prescribe the presence of a non-hyperbolic equilibrium, $\bx_{\alpha_p}$, and the existence of a parameter dependent, one-dimensional, attracting manifold near $(\bx_{\alpha_p},\alpha_p)$ for which \eqref{eq:bifode1} is l.t.e.\ to a system
\begin{equation}\label{eq:genfold}
        \left\{
    \begin{aligned}
        \dot{u} &=  f(u,\alpha), \\
        \dot{\bv} &= \bA(u,\alpha)\bv,\\
    \end{aligned}
    \right.
\end{equation}
where for each $t\ge 0$, $u(t)\in\bbR, \ \bv(t)\in\bbR^{n-1}, \ f:\bbR\times\bbR\to\bbR$ is smooth and $\bA(u(t),\alpha)$ is an $(n-1)\times (n-1)$ matrix with all of its eigenvalues having negative real part. 
For the generic fold bifurcation system \eqref{eq:genfold}, the non-hyperbolic equilibrium present at $\alpha_p$ disappears for $\alpha<\alpha_p$ while two equilibrium points, one stable and one unstable, emerge for $\alpha>\alpha_p$. 
Therefore, a system satisfying the conditions (B.I-III) presents two stable and one unstable equilibrium points for $\alpha>\alpha_p$.  
That is, we arrive at the following result.

\begin{remark}[Bi-stability]\label{rem:bistability}
For a system as in \eqref{eq:bifode1} satisfying the conditions \ref{BI}-\ref{BIII}, with value of $\alpha$ such that $\alpha - \alpha_p$ is positive and sufficiently small, there exist $\bx^{(1)}_*,\bx_\alpha,\bx^{(2)}_*\in \bbR^n$ with $\bF(\bx^{(1)}_*,\alpha)=\bF(\bx_\alpha,\alpha)=\bF(\bx^{(2)}_*,\alpha)=\bzero$ and all the eigenvalues of the Jacobian matrices $\bJ(\bx^{(1)}_*,\alpha)$ and $\bJ(\bx^{(2)}_*,\alpha)$ are negative (linear stability), while $\bJ(\bx_\alpha,\alpha)$ has one positive eigenvalue  (instability) and the rest are negative.
\end{remark}

Similarly to Hopf bifurcation systems in Section \ref{sec:hopf}, the image of the LNA, $\bX(t) = \bx(t) +\Omega^{-1/2}\bxi(t)$ under the local homeomorphism reveals stochastic processes $U(t)\approx u(t) + \Omega^{-1/2}\ \xi_u(t)$, evolving in the single centre coordinate; and $\bV(t)\approx \bv(t) + \Omega^{-1/2}\bm \xi_\bv(t)$, evolving in the stable coordinates. Their stochastic perturbations, $\xi_u(t)$ and $\bm \xi_\bv(t)$, evolve according to \eqref{eq:eigenSDE}.

\subsection{The pcLNA algorithm for bi-stable systems}\label{sec:phasebi-stable}

A system satisfying Remark \ref{rem:bistability} with $\alpha>\alpha_p$ exhibits two stable equilibrium points. 
The standard LNA, which relies on a single deterministic solution to simulate a trajectory, will produce paths centered around the ``nearest'' equilibrium. As a result, it will fail to approximate trajectories produced by the SSA whenever there is a substantial probability that a stochastic trajectory produced by the SSA escapes from this equilibrium (see Figure \ref{fig:LNA}(D) and SM Figure~\ref{sifig:toggleswitchLNA}). 

Here we discuss how to choose a metric $d$ and a set of reference trajectories $\Gamma$ (following the procedure in \ref{sec:phaseCorrConseq}) in order to apply the pcLNA algorithm in \ref{alg:gen_pclna} and obtain accurate simulations. 

If $\bR_c=\bu_1$ is the eigenvector corresponding to the zero eigenvalue of $\bJ(\bx_{\alpha_p},\alpha_p)$, then we define the projection map $\pr:\bbR^n\to \bbR$, onto the eigenspace that assigns to each state $\bx$ its approximate location in the centre coordinate: $u=\pr(\bx)=\bm P_c(\bx - \bx_{\alpha_p})$ where $\bm P_c:=({\bR_c}^\top\bR_c)^{-1}{\bR_c}^\top$. 
Again, the distance metric in \eqref{eq:dmetric} 
is
$ d(\bX,\bx) = |U - u|, $
where $U=\pr(\bX)$ and $u=\pr(\bx)$. 

When selecting the set of reference trajectories $\Gamma$, at least two trajectories—each converging to a different equilibrium—are required.
This is necessary to allow stochastic trajectories that start from the same initial condition to reach the vicinity of different equilibria via LNA transitions, a behaviour that is observed in SSA trajectories.

Having defined the distance metric $d$ and the set of reference trajectories $\Gamma$, we can apply \eqref{eq:phaseargmin} and \eqref{eq:phasecorrect} to define the phase-correcting map $\bG$ and therefore apply the Algorithm~\ref{alg:gen_pclna}. 

With this in place, we now present a pcLNA algorithm specifically for bi-stable systems in Algorithm~\ref{alg:bi-stable_pclna}.
In Section~\ref{sec:numerical_investig}, we show that this simple phase map is sufficient to achieve high levels of accuracy. 

\begin{algorithm}[h!]
\caption{pcLNA for bi-stable systems}\label{alg:bi-stable_pclna}
\textbf{Inputs:}\\
a.\ Initial conditions: $s_0=0=t_0$, $j_0\in \{1,2\}$, $\bX(0)$, $\{\bx^{(j)}(0) :   j=1,2\}$. Parameters: $\Delta t$, $\Omega$.\\ 
b.\ System satisfying Remark \ref{rem:bistability} with solutions $\gamma_j=\{\bx^{(j)}(t) :   t\geq 0\}$ converging to $\bx_j^*$, $j=1,2$, respectively, and their corresponding solutions $\bC^{(j)}$ and $\bD^{(j)}$ of \eqref{eq:Code}, \eqref{eq:Vode}, respectively.\\
c. The projection map $\pr:\bbR^n\to \bbR$, $u=\pr(\bx)=\bm P_c(\bx - \bx_{\alpha_p})$ where $\bm P_c = \bu^\top$ with $\bu$ a normalised eigenvector corresponding to the zero eigenvalue of $\bJ(\bx_{\alpha_p},\alpha_p)$.

\textbf{Steps:}
Compute $\{u^{(j)}(t):t\geq 0\}$ where $u^{(j)}(t)=\pr(\bx^{(j)}(t))$ for $j=1,2$, and $\bkappa(s_0)=\Omega^{1/2}(\bX(0)-\bx^{(j_0)}(0))$. For $i=1,2,\dots$
\begin{enumerate}[label=\arabic*.]
    \item draw sample $\bxi(s_{i-1}+\Delta t)$ from the MVN distribution with mean $\bC^{(j_{i-1})}(s_{i-1},s_{i-1}+\Delta t)\bkappa(s_{i-1})$, and covariance matrix $\bD^{(j_{i-1})}(s_{i-1},s_{i-1}+\Delta t)$;
    \item compute $\bX(t_i) = \bx^{(j_{i-1})}(s_{i-1}+\Delta t) + \Omega^{-1/2}\bxi(s_{i-1}+\Delta t)$;
    \item compute $U = \pr(\bX(t_i))$ and find $(j_i,s_i)=\argminB_{\ell\in \{1,2\},t\geq 0} \| U-u^{(\ell)}(t)\|$;
    \item compute $\bkappa(s_i) = \Omega^{1/2}(\bX(t_i)-\bx^{(j_i)}(s_i))$.
\end{enumerate}
\textbf{Outputs:} $\{\bX(t_i) :   i=0,1,2,\dots \}$.
\end{algorithm}

\begin{figure}[h]
    \centering
    \includegraphics[width=\textwidth]{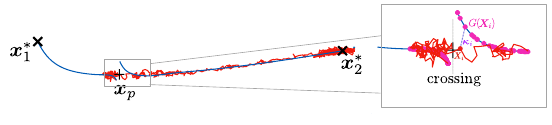}
    \caption{\label{fig:PCbi-stable} Phase Correction (PC) for bi-stable systems. (Left) The stochastic trajectory (red) and two deterministic trajectories (blue), each converging to a different stable equilibrium (x) on the two sides of the unstable equilibrium (+) are displayed. (Right) Zoom in to the part of simulation where a crossing from one side of the unstable equilibrium to the other was performed with the PC mapping $\bG$ (pink circles) and the deviation $\kappa_i$ shown. 
    }
\end{figure}

\section{Numerical investigations}\label{sec:numerical_investig}

As mentioned in Section \ref{sec:LNA}, the LNA boasts fast computation of distributions of the stochastic process $\{\bX(t) :   t\geq 0\}$ as it only demands solving a set of ODEs to derive the RRE solutions $\{\bx^{(j)}(t) :   t\geq 0\}$, $j=1,\dots,J$ in \eqref{eq:bifode1} and the corresponding drift and diffusion matrix solutions, $\bC^{(j)}$ and $\bD^{(j)}$, of Eq.\ (\ref{eq:Code}-\ref{eq:Vode}). 
Since the pcLNA algorithm also only requires such solutions, it also inherits its speed of simulation. 

Next, we study the accuracy and computational efficiency of the pcLNA compared to the SSA. 
For this, we separately use SSA and pcLNA to produce a large number of long-time, stochastic trajectories and then compare their empirical probability distributions at various time-points. 
We perform this exercise for various systems, either presenting bi-stability or a Hopf bifurcation. 
All numerical investigations were performed on a MacBook Pro workstation with an Apple M2 processor (8 CPU cores) and 16 GB RAM using the software PeTTSy \cite{Domijan2016} in MATLAB \cite{MATLAB}.

For Hopf bifurcation systems, we perform comparisons for all three qualitatively different cases where the system's complex conjugate pair of eigenvalues in \ref{HII} has $\mu(\alpha)<0,=0,>0$. 
In these three cases, the limiting RRE system presents quickly damped oscillations with a focus shape, slowly damped oscillations, and sustained oscillations, respectively (see also Figure \ref{fig:dynamics}A and \ref{fig:hopf}).
For bi-stable systems, we focus on parameter values satisfying \ref{rem:bistability} (see Figure \ref{fig:dynamics}B, right) where the system presents bi-stability. 

\begin{figure}[h]
    \centering
    \includegraphics[width=0.7\textwidth]{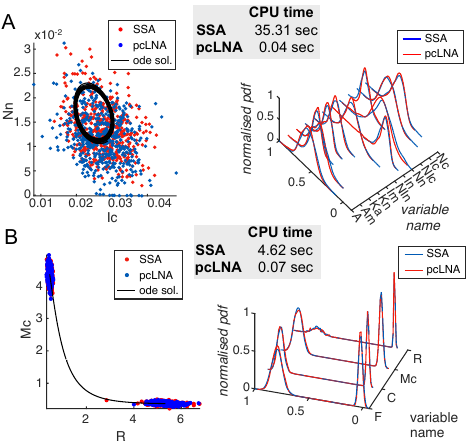}
    \caption{\label{fig:Numerical} Comparison between SSA (red) and pcLNA (blue) for (A) the NF-$\kappa$B \cite{ashall} and (B) the somitogenesis switch \cite{Goldbeter2007} system.   
    (Left) The states recorded after four cycles (A) and at $t=100$ min (B) of respectively $n=500$ and $n=1000$ simulations of the two systems and solutions of the corresponding RRE in \eqref{eq:bifode1}.
    (Right) The corresponding empirical probability density functions of each of the system variables. 
    The median CPU times for deriving one stochastic trajectory are also reported. See SM Section~\ref{SIsec:numerics}
    for the details of the simulations.
    }
\end{figure}

The Hopf bifurcation systems used for comparisons are the following.
\begin{itemize}
    \item The Brusselator system in \cite{LEFEVER1971267}, which is a classical oscillatory system involving complex interactions between two variables (for more details, see SM Section~\ref{sisec:brus}).
    \item The three-variable system in \cite{WILHELM1995} that is the smallest reaction network presenting a Hopf bifurcation (see SM Section~\ref{sisec:simpleHopf}). 
    \item The response of the \NFKB signalling system to \TNFA signals in \cite{ashall}, which is a system that involves 11 molecular populations and 24 reactions (see SM Section~\ref{sisec:nfkb}). 
\end{itemize}
For the first two systems above, we produced $1000$ stochastic trajectories using SSA and pcLNA for a time interval of length $8\tau$, where $\tau$ is the length of one period of the corresponding system. 
For the third system (where SSA simulations are substantially slower), we perform $500$ repetitions for a time-length of $4$ periods of the system.
We recorded the time taken to compute one stochastic trajectory for these time intervals, and then computed the median of these times, which we call CPU time in all figures.
We also collected using these simulations the state vector $(X_1(t_j), \dots, X_n(t_j))^\top$ at various time-points, $t_j$, $j=1,2,\dots,T$ in each repetition, and computed the corresponding empirical density functions for the probability distribution of each variable $X_i(t_j)$ using kernel density estimation in MATLAB \cite{MATLAB}.
We present the results in Figure \ref{fig:Numerical}A and SM Figures 
\ref{sifig:nfkb015}-
\ref{sifig:simpleHopf7}.

We clearly see that the pcLNA and SSA distributions are nearly indistinguishable, while the CPU time is reduced by a magnitude of order $O(10)-O(10^3)$. 
The reduction in CPU time depends on the complexity of the system (number and speed of reactions) and the system size. 
Larger system sizes imply larger number of reaction occurrences and slower SSA simulations, while CPU times for pcLNA are not affected by system size.

The bi-stable systems considered are the following.
\begin{itemize}
    \item The genetic toggle-switch system in \cite{Gardner2000}, which involves two variables (for more details, see SM Section~\ref{sisec:toggleswitch}).
    \item The two-variable, simplified cell-cycle system presented in \cite{Angeli2004} (see SM Section~\ref{sisec:cellCycle}).
    \item The somitogenesis switch system, presented in \cite{Goldbeter2007}, that involves four variables and ten reactions (see SM Section~\ref{sisec:somito}).
\end{itemize}

We produced $1000$ trajectories using SSA and pcLNA for the above systems, running for time intervals of length sufficient for convergence to one of the two stable equilibria by the solution of the corresponding RRE. 

Note that if we initialise the system at a point close to one of the stable equilibria, then nearly all trajectories converge to the same equilibrium and the SSA simulations are well approximated by the standard LNA. 
Instead, we focused on simulations where the initial conditions are close to the unstable equilibrium so that there is a substantial probability of convergence of a given stochastic trajectory for both equilibria. 
In this case, all the standard LNA trajectories will converge to the same equilibrium and therefore can fail badly to approximate the SSA simulations (see Figure \ref{fig:LNA}(D) and SM Figure~\ref{sifig:toggleswitchLNA}). 

The pcLNA algorithm \ref{alg:bi-stable_pclna} requires the choice of at least two solutions of the RRE that differ only by their initial condition and converge to different equilibria. 
In the simulations presented, the two initial conditions are of the form $(1-\epsilon)\,\bx_\alpha + \epsilon\,\bx^{(j)}_* + \bc$, $j=1,2$. Here $0\leq \epsilon \leq 0.1$, $\bx_\alpha$ is the unstable equilibrium and $\bx^{(j)}_*$, $j=1,2$, the stable equilibria. 
The real vector $\bc$ performs a change of location in a direction orthogonal to the eigenvector corresponding to the real positive eigenvalue of the Jacobian $\bJ(\bx_\alpha)$.
This change of location reveals some short-time transient dynamics, but it is not necessary if one is interested in the long-time dynamics. 
For the specific initial conditions and other details, see SM Sections~\ref{sisec:toggleswitch}-\ref{sisec:somito}.

The results are presented in Figure~\ref{fig:Numerical}B and SM Figures~\ref{sifig:toggleswitch}, \ref{sifig:cellcycle},
\ref{sifig:SI_devGoldbeter_omega500}. 
The two methods, SSA and pcLNA, are compared in the same way as the Hopf bifurcation system and the results again show the agreement between the two methods. Again, there are significant reductions in the CPU time, especially for the Somitogenesis switch system (where pcLNA CPU time is $\approx 58$ times smaller). 

Finally, to quantify the computational cost of SSA and pcLNA across systems, we report the CPU times in Table~\ref{tab:times}. We also introduce a simple and practical predictor of the SSA runtime based on the system’s average reaction activity.

For a given system, the total propensity
$
\pi_0(\bY)=\sum_{j=1}^{r}\pi_j(\bY),
$
is evaluated along the deterministic trajectories $\{\bx^{(\ell)}(t):t\in[0,T],\ell\in \cL\}$ used for pcLNA simulation taking $\bY=\Omega\,\bx^{(\ell)}(t)$ (here $\cL=\{1,\dots,L\},\, L\in\{1,2\}$). 
This yields the time-dependent reaction activity of the mean-field dynamics which, although it neglects stochastic deviations, provides a reliable summary of the system’s reaction intensity over different dynamical phases.
We then define the average total propensity
\[
\bar{\pi}_0=\int_0^T \pi_0(\Omega\bx(t))\,dt.
\]
Since the SSA explicitly simulates every reaction event, and each event has constant computational cost, the expected number of reactions in time $T$, given by $N_r=T\bar{\pi}_0$, provides an easily-computable, interpretable predictor of SSA runtime.

For the systems in Table~\ref{tab:times}, this predictor correlates extremely strongly with the measured SSA CPU times ($\rho=0.9963$).
In contrast, we see that systems with similar number of reactions or number of populations have CPU times of different orders of magnitude for both SSA and pcLNA. 
In the same table, we see that pcLNA runtime compared to SSA is orders of magnitude smaller and grows much more slowly with system complexity. For example, while SSA is $O(10^4)$ times slower for the NF-$\kappa$B model than for the toggle-switch system, pcLNA times are of the same order.

\begin{table}[!ht]
\centering
\begin{tabular}{llllll}
\toprule
& & & & \textbf{CPU time} & \\
System & $N_r/10^3$ & $n$ & $r$ & SSA & pcLNA\\
\midrule
toggle-switch \cite{Gardner2000} & 5.0347 & 2 & 4 & 0.0296 & 0.0110\\
cell-cycle \cite{Angeli2004} & 4.4595 & 2 & 6 & 0.0318 & 0.0065\\
somitogenesis-switch \cite{Goldbeter2007} & 821.18 & 4 & 10 & 4.6188 & 0.0690\\
smallest Hopf \cite{WILHELM1995} (focus) & 376.187 & 3 & 5 & 2.7065 & 0.0184\\
smallest Hopf \cite{WILHELM1995} (Hopf) & 468.307 & 3 & 5 & 3.3828 & 0.0181\\
smallest Hopf \cite{WILHELM1995} (limit-cycle) & 564.457 & 3 & 5 & 3.9284 & 0.0173\\
brusselator \cite{LEFEVER1971267} (focus) & 360.305 & 2 & 4 & 3.4655 & 0.0196\\
brusselator \cite{LEFEVER1971267} (Hopf) & 405.078 & 2 & 4 & 3.8119 & 0.0171\\
brusselator \cite{LEFEVER1971267} (limit-cycle) & 550.039 & 2 & 4 & 3.2390 & 0.0417\\
NF-$\kappa$B \cite{ashall} (focus) & 10986.9 & 11 & 25 & 44.8228 & 0.0276\\
NF-$\kappa$B \cite{ashall} (Hopf) & 9160.79 & 11 & 25 & 35.3062 & 0.0390\\
NF-$\kappa$B \cite{ashall} (limit-cycle) & 7661.71 & 11 & 25 & 27.1758 & 0.0264\\
\bottomrule
\end{tabular}
\caption{Computational time of SSA and pcLNA for different systems. The median CPU times for the SSA and pcLNA simulations with parameter values given in the SM Sections~\ref{sisec:nfkb}-\ref{sisec:somito} are displayed along with the number of populations ($n$), reaction channels ($r$), and average number of reaction occurrences in thousands ($N_r$, see \ref{sec:numerical_investig}).}\label{tab:times}
\end{table}

\section{Discussion}\label{discussion}

The results in this paper demonstrate the capability of the modified LNA method to effectively capture non-linear dynamics over long time intervals. The modification introduced is justified by a novel reformulation of the LNA in coordinates that separate persistent non-linear dynamics from transient dynamics. The accuracy of the approximation, combined with the analytical tractability of the LNA method and speed of computation, establishes a robust framework for investigating population dynamics, and fitting these models to data. It overcomes key limitations of existing methods and opens new avenues for advancing the stochastic modelling of dynamical systems.

We showcase the versatility of our methodology across a range of systems exhibiting two fundamental forms of non-linear behaviour—oscillations and bi-stability—which are common in biology and other scientific domains. 
Beyond these cases, our approach, grounded in centre manifold theory from dynamical systems, is applicable to a much broader class of reaction networks. Specifically, for any parameter-dependent network whose RRE solution features a non-hyperbolic equilibrium—a scenario frequently encountered in practice—the pcLNA algorithm enables fast and accurate simulation of long-term stochastic trajectories, even in large and complex systems.

Such stochastic simulations are essential for applications including in silico experimentation, and sensitivity analysis \cite{Komorowski2011,Gupta2014d,Vanlier2014,Rainforth2024,Liepe2014,Shahmohammadi2020,Chhajer2024,Minas2019a}, and statistical inference using Approximate Bayesian Computation methods \cite{Tavar1997,Marjoram2003,Sisson2007,Toni2009}. Moreover, this work lays the groundwork for future methods that leverage the long-term accuracy and analytical strengths of pcLNA—enabling, for instance, the exact computation of likelihood functions for time-series data, as well as key quantities like the Fisher information, which are critical for assessing system sensitivity and robustness to parameter variation.

\bibliographystyle{plain}
\bibliography{main}

\newpage
\newpage

\appendix

\renewcommand\thesection{S\arabic{section}}
\renewcommand\thetable{S\arabic{table}}
\renewcommand\thefigure{S\arabic{figure}}

\begin{center}
    \Huge \textbf{Supplementary Materials (SM)}
\end{center}
\vspace{1em}   

The Supplementary Materials (SM) contain further mathematical details and numerical investigations referenced in the main paper.
We used PeTTSy \cite{Domijan2016} implemented in the MATLAB \cite{MATLAB} environment and available at \url{https://wrap.warwick.ac.uk/id/eprint/77654/} to produce all results.

\section{The stochastic simulation algorithm (SSA)}\label{app:SSA}

Consider paths of $\{\bY(t) \ | \ t\geq 0\}$ whose time-evolution is given by
the random time change representation (RTC),
\begin{equation}\label{eq:rtc2}
	\bY(t) = \bY(0) + \sum_{j=1}^r \bnu_j N_j \left( \int_0^t \pi_j(\bY(s))ds  \right),
\end{equation} 
where, for $j\in\{1,\dots,r\}$, $N_j$ are independent, unit-rate Poisson processes corresponding to reaction $R_j$.
% we do not look to the solution $P(\by,t)$ of the chemical master equation \eqref{eq:master}. Instead 
To simulate such paths, the stochastic simulation algorithm (SSA) generates the next reaction from any given state. 
That is, if $\hat{\uptau} > 0$ and $k \in \{1,2,\dots,r \}$ correspond to the time to the next reaction and the index of the next reaction, respectively,
it can be shown that, given the current state at time $t\geq 0$, $\bY(t) = y$, the joint probability density function $p(\hat{\uptau},k \ | \ \by,t)$ of $\hat{\uptau}$ and $k$ is
\begin{equation}\label{eq:SSAprob}
    p(\hat{\uptau},k \ | \ \by,t) = \pi_{k}(\by) \ \exp{\left(-\sum_{j=1}^r \pi_{j}(\by)\hat{\uptau}\right)}.
\end{equation}
where $\pi_{k}$ is the propensity function of reaction $k$. 
Informally, $p(\hat{\uptau},k  \ | \ \by,t)d\hat{\uptau}$ represents the probability that, given $\bY(t) = \by$, the next reaction in the system will occur in the infinitesimal time interval $[t+\hat{\uptau},t+\hat{\uptau}+d\hat{\uptau})$ and will be a $R_{k}$ reaction. 
Equation \eqref{eq:SSAprob} implies that $\hat{\uptau}$ is an exponential random variable with mean $1/\sum_{j=1}^r \pi_{j}(\by)$ and $k$ is an integer random variable with point probabilities $\pi_{k}(\by)/\sum_{j=1}^r \pi_{j}(\by)$ and statistically independent of $\hat{\uptau}$. 
Hence we can sample such random variables exactly using the following; generate two random numbers, $r_1$ and $r_2$, from the uniform distribution in the unit interval and take,
\begin{equation}\label{eq:tausample}
    \hat{\uptau} = -\frac{\log(r_1)}{\sum_{j=1}^r \pi_{j}(\by)},
\end{equation}
and $k$ to be the smallest integer with $k\leq r$ such that,
\begin{equation}\label{eq:jsample}
    \sum_{j=1}^{k} \pi_{j}(\by) > r_2 \sum_{j=1}^r \pi_{j}(\by).
\end{equation}
Using the sampling procedure given by \eqref{eq:tausample} and \eqref{eq:jsample}, we are able to construct the SSA for paths of $\{\bY(t) \ | \ t\geq 0\}$ which evolve exactly according to \eqref{eq:rtc2}. This is written in Algorithm \ref{alg:ssa}.

Here the output sequence $(t_j,\by_j)$, $j=1,2,\dots$, fully describes the evolution of a stochastic trajectory $\{\bY(t):0\leq t\leq T\}$. Note that $\bY(t)=\by_j$ for $t_{j}\leq t<t_{j+1}$.
These paths are exact consequences of the chemical physics of stochastic reaction networks and hence capture the true dynamics of the network. For further details see \cite{Gill76,Gill77}. 
Consequently, one can compare distributions of states simulated by a model to those simulated by the SSA in order to test the statistical accuracy of the model. 

\begin{algorithm}
\caption{Stochastic Simulation Algorithm (SSA)}\label{alg:ssa}
\textbf{Inputs:}
Initial conditions: $t_0=0$, $\bY(0)=\by_0$, $i=0$, parameter $T>0$. Initialise $t\leftarrow t_0$ and $\by\leftarrow \by_0$.\\ 
\textbf{Steps:}
While $t<T$
\begin{enumerate}
    \item generate two random numbers, $r_1$ and $r_2$, from the uniform distribution on the open unit interval;
    \item sample $\hat{\uptau}$ according to \eqref{eq:tausample} and $k$ according to \eqref{eq:jsample};
    \item update $t \leftarrow t+\hat{\uptau}$ and $\by \leftarrow \by+\bnu_k$;
    \item update $i \leftarrow i+1$ and store $t$ in $t_i$ and $\by$ in $\by_i$.
\end{enumerate}
\textbf{Outputs:} $(t_j,\by_j)$, $j=1,\dots,i$
\end{algorithm}
% The result is the numerical simulation $\{\bY(t_i) \ | \ i\in\bbNz\}$ where $(t_i)_{i\in\bbNz}$ is a sequence of times, almost certainly not equally spaced. 
% Note that 

When performing the SSA simulations, we used the so-called \emph{thinning} method to reduce the computational memory used. That is, we used Algorithm~\ref{alg:ssa_thinning}.
\begin{algorithm}[h!]
\caption{SSA with thinning}\label{alg:ssa_thinning}
\textbf{Inputs:}
Initial conditions: $t_0=0$, $\bY(0)=\by_0$, $i=0$, $m=1$. Parameter $T>0$, $\delta t>0$, $M$ positive integer.\\ 
\textbf{Steps:}
While $t<T$ and $m\leq M$
\begin{enumerate}
    \item generate two random numbers, $r_1$ and $r_2$, from the uniform distribution on the unit interval;
    \item sample $\hat{\uptau}$ according to \eqref{eq:tausample} and $k$ according to \eqref{eq:jsample};
    \item update $t \leftarrow t+\hat{\uptau}$ and $\by \leftarrow \by+\bnu_k$;
    \item update $i \leftarrow i+1$ and
        \subitem  \textit{if} $t \in (m \delta t, (m+1)\delta t)$, then store $t$ in $t_m$ and $\by$ in $\by_m$, and update $m \leftarrow m+1$, 
        \subitem \textit{else if} $t > (m+1)\delta t$, end procedure with message ``increase $\delta t$ and run again.''.
\end{enumerate}
\textbf{Outputs:} $(t_j,\by_j)$, $j=1,\dots,m$
\end{algorithm}
This records the SSA generated states at the first reaction times after each $m\delta t$, $m=0,1,2,\dots,M$, for some large $M$ and fairly small $\delta t$. 
Furthermore, when we wish to compare SSA with pcLNA at specific times we use a smoothing spline interpolation to get the SSA state (and similarly pcLNA) at a specific time, say $t \in [(m-1)\delta t,m\delta t]$.

\section{The derivation of the Linear Noise Approximation (LNA)}
\label{app:lna}

The following is a heuristic derivation of the LNA equations, similar to that provided in \cite{Anderson2011}. For a formal derivation see \cite{KURTZ1978223}.
To simplify notation, in this section we use the notation $x_t = x(t)$, for all states considered (i.e.\ $\bx$, $\bxi$, $\bX$). 
The LNA ansatz,
\begin{equation*}\label{SIeq:lnaansatzX}
    \bX_t = \bx_t + \Omega^{-1/2}\bxi_t,
\end{equation*}
implies that 
$$ \bxi_t = \sqrt{\Omega}(\bX_t - \bx_t) $$
and thus
$$ \bxi_{t+dt} - \bxi_t = \sqrt{\Omega}\left( (\bX_{t+dt} - \bX_t) - (\bx_{t+dt} - \bx_t) \right).  $$
Using the form of RTC for $\bX$ and the RRE, we get that
$$ \bxi_{t+dt} - \bxi_t = \sqrt{\Omega}\left(\sum_{j=1}^r \bnu_j \left\{ \Omega^{-1} N_j\left(\Omega \rho_j( \bX_t) dt \right)  - \rho_j( \bx_t) dt \right\} \right).$$
We next insert in the above sum the zero terms $0=\rho_j( \bX_t)dt -\rho_j( \bX_t)dt$, where $\rho_j( \bX_t)dt$ is the mean of $\Omega^{-1} N_j\left(\Omega \rho_j( \bX_t) dt \right)$, to get 

    \begin{equation*}
        \bxi_{t+dt} - \bxi_t = \sqrt{\Omega}\left(\sum_{j=1}^r \bnu_j \left\{ \Omega^{-1} N_j\left(\Omega \rho_j( \bX_t) dt \right) - \rho_j( \bX_t)dt + \left(\rho_j( \bX_t) - \rho_j( \bx_t) \right) dt \right\} \right).
    \end{equation*}
Then, by the Central Limit Theorem, as $\Omega \to \infty$, 
$$ Z = \frac{\Omega^{-1}N_j\left(\Omega \rho_j( \bX_t) dt \right) - \rho_j( \bX_t)dt }{ \sqrt{\Omega^{-1} \rho_j( \bX_t) dt } } \sim N(0,1) $$
where $N(0,1)$ is the standard normal distribution. 
Then for sufficiently large $\Omega$, 
\begin{equation}\label{eq:LNAclt}
 \sqrt{\Omega}\left(\Omega^{-1}N_j\left(\Omega \rho_j( \bX_t) dt \right) - \rho_j( \bX_t)dt \right) \approx Z\sqrt{\rho_j( \bx_t) dt }. 
 \end{equation}
Furthermore, by applying a Taylor expansion of $\rho_j( \bX_t)$ about $\bx_t$, we get 
\begin{equation}
    \sqrt{\Omega} \left( \rho_j( \bX_t) - \rho_j( \bx_t) \right) = \nabla_{\!\bx}\, \rho_j(\bx_t)^\top\bxi_t + O\left(\Omega^{-1/2}\right),
\end{equation}
where $\nabla_{\!\bx}^\top = (\partial/\partial x_1, \dots,  \partial/\partial x_n)$.     
Hence, for sufficiently large $\Omega$,
$$ \bxi_{t+dt} - \bxi_t \approx  \sum_j \bnu_j \bxi(t)^\top\nabla_{\!\bx}\, \rho_j(\bx_t) dt + \sum_j Z_j \bnu_j \sqrt{\rho_j( \bx_t)dt}  $$
where $Z_j$ are independent random variables with standard normal distribution.
Writing the above equation in matrix form, we get 
\begin{equation}\label{eq:langevin2}
    d\bxi_t = \bJ_t\bxi_t dt + \bE_t\, d\!\bB_t,
\end{equation}
where $\bJ_t=\bJ(\bx_t)$ is the Jacobian matrix of the RRE \eqref{eq:bifode1}, and 
$$\bE_t=\bE(\bx_t)=\bS\text{diag}\left(\!\sqrt{\rho_1(\bx_t)},\dots,\sqrt{\rho_r(\bx_t)}\right)$$ where
$\bS=[\bnu_1 \cdots \bnu_r]$ is known as the stoichiometry matrix, and $\bB_t$ is an $r$-dimensional Wiener process.

\section{Centre manifold theory} \label{SIsec:man}
For a more detailed introduction to centre manifold theory see \cite{Kuznetsov2011}.

Different dynamical systems can be grouped together based on the qualitative similarity of their behavior near their equilibrium points. 

\begin{definition}[Topological equivalence]\label{def:topequiv}
Consider two ODE systems as in \ref{eq:bifode1} with velocities $\bF$ and $\bF'$ on $\bbR^n$ and $\bbR^m$ respectively. 
Suppose they each have an equilibrium point $\bx_0$ and $\bx_0'$, respectively. 
Then, near their equilibria, the two dynamical systems are \emph{locally topologically equivalent} if there exists a homeomorphism $\bh:\bbR^n\to\bbR^m$ that is
\begin{enumerate}[label=(\roman*)]
    \item defined in a neighbourhood $A\subset\bbR^n$ of $\bx_0$;
    \item satisfies $\bx_0'=\bh(\bx_0)$;
    \item maps paths of the system with velocity $\bF$ in $A$ onto paths of the system with velocity $\bF'$ in $\bh(A)\subset\bbR^m$, preserving the direction of time.
\end{enumerate}
\end{definition}

Intuitively, this means the paths of one dynamical system near its equilibrium can be continuously deformed onto paths of the other dynamical system near its equilibrium, and hence both systems exhibit qualitatively similar behavior near their equilibria.

\paragraph{Extended system and non-hyperbolic equilibrium}
Consider the deterministic evolution solving the reaction rate equation \eqref{eq:bifode1} when one parameter $\alpha\in\bbR$ is free to vary. 
Assume there exists $\alpha_p\in\bbR$ such that
\begin{equation}\label{eq:nonhyper}
 \bF(\bx_{\alpha_p},\alpha_p)=\bm 0 \quad \text{and} \quad n_0>0,
\end{equation}
where $n_0,n_+,n_-$ denote the number of eigenvalues of the Jacobian matrix $\bJ(\bx_{\alpha_p},\alpha_p)$ with zero, positive and negative real part, respectively. 

Introduce the extended autonomous system
\begin{equation}\label{bifode2}
\left\{
\begin{aligned}
\dot{\bx} &= \bF(\bx,\alpha), \\
\dot{\alpha} &= 0,
\end{aligned}
\right.
\end{equation}
whose Jacobian at $(\bx,\alpha)=(\bx_{\alpha_p},\alpha_p)$ is
\begin{equation}\label{extjac}
\mathcal A :=
\begin{pmatrix}
\bJ(\bx_{\alpha_p},\alpha_p) & \partial_\alpha \bF(\bx_{\alpha_p},\alpha_p) \\
0 & 0
\end{pmatrix}.
\end{equation}
This matrix has $n_0+1$ eigenvalues with zero real part and $n-n_0$ eigenvalues with nonzero real part.

\begin{theorem}[Centre Manifold Theorem]\label{thm:man}
Let $E^c$ denote the eigenspace of $\mathcal A$ corresponding to the $n_0+1$ eigenvalues with zero real part. 
Then there exists a smooth $(n_0+1)$-dimensional invariant manifold $\mathcal W^c$, defined locally near $(\bx_{\alpha_p},\alpha_p)$ and tangent to $E^c$ at that point.

Since $\dot\alpha=0$, the hyperplanes
\[
\Pi_{\alpha'} := \{(\bx,\alpha)\mid \alpha=\alpha'\}, \qquad \alpha'\in\bbR,
\]
are invariant under \eqref{bifode2}. Consequently, for each $\alpha$ near $\alpha_p$, the centre manifold is foliated by $n_0$-dimensional invariant manifolds
\[
\mathcal W^c_\alpha := \mathcal W^c \cap \Pi_\alpha ,
\]
which we call the \emph{centre manifolds}. Moreover, for each such $\alpha$, there exists a neighbourhood $\Delta(\bx_{\alpha_p})$ such that any trajectory of \eqref{bifode2} starting in $\Delta(\bx_{\alpha_p})$ either converges to $\mathcal W^c_\alpha$ or leaves $\Delta(\bx_{\alpha_p})$ exponentially fast as $t\to\infty$. 
\end{theorem}

\paragraph{Takens normal form for the extended system}
Let $\lambda_1,\dots,\lambda_{n_{cu}}$, where $n_{cu}=n_0+n_-+1$, denote the eigenvalues of $\mathcal A$ with zero or negative real part including the zero eigenvalue associated with $\dot\alpha=0$. For each integer $k\geq 1$, we say that \eqref{bifode2} satisfies the \emph{$k$-Sternberg nonresonance condition} 
(see \cite[Definition 2 and Section 4]{Takens1971})
if the following holds:
for any choice of integers 
% $c_1,\dots,c_{n_0+n_-+1}$ 
$c_1,\dots,c_{n_{cu}}$ 
satisfying
\[
2 \le \sum_{i=1}^{n_{cu}} c_i \le \gamma(\mathcal A,k),
\]
% \[
% 2 \le \sum_{i=1}^{n_0+n_-+1} c_i \le \gamma(\mathcal A,k),
% \]
the sums,
\begin{equation*}
\sum_{i=1}^{
% n_0+n_-+1
n_{cu}
} c_i\,\lambda_i \neq 0,
\qquad
-\lambda_j + \sum_{i=1}^{
% n_0+n_-+1
n_{cu}
} c_i\,\lambda_i \neq 0
\quad \text{for all } j=1,\dots,
% n_0+n_-+1
n_{cu}.
\end{equation*}

Here the integer $\gamma(\mathcal A,k)$ is defined as follows. 
Here we use the notation $\Re(\mu)$ to denote the real part of an eigenvalue $\mu$.
Let $\mu_1,\dots,\mu_{n_-+n_+}$ denote the eigenvalues of $\mathcal{A}$ with non zero real part, 
ordered so that
\[
\Re(\mu_1)\le \cdots \le \Re(\mu_{n_-})<0<\Re(\mu_{n_-+1})\le \cdots \le \Re(\mu_{n_-+n_+}).
\]
Define
\[
\overline{M}:=\max_{i>n_-}\Re(\mu_i), \qquad
\overline{m}:=\min_{i>n_-}\Re(\mu_i), \qquad
\overline{N}:=\max_{i\le n_-}|\Re(\mu_i)|, \qquad
\overline{n}:=\min_{i\le n_-}|\Re(\mu_i)|.
\]
Then $\beta(\mathcal A,k)$ is the smallest integer such that
\[
\overline{N} + r\,\overline{M} - (r-\beta(\mathcal A,k))\,\overline{n} < 0
\qquad \text{for all integers } r \le k,
\]
and $\gamma(\mathcal A,k)$ is the smallest integer such that
\[
\overline{M} + r\,\overline{N} - (r-\gamma(\mathcal A,k))\,\overline{m} < 0
\qquad \text{for all integers } r \le \beta(\mathcal A,k).
\]

The above conditions hold for almost all values of parameter $\alpha$.  
% except on lower-dimensional subsets of parameter space. 
In reaction networks, the free parameters $\alpha$ are 
typically reaction rate constants, 
for which the eigenvalues $\lambda_i$ of the Jacobian depend smoothly. 
Resonance requires exact relations of the form $\sum_i c_i \lambda_i = 0$ for integers $c_i$, which occur only on sets of Lebesgue measure zero in parameter space and are destroyed by arbitrarily small parameter perturbations. 
Consequently, for analytical purposes, the nonresonance conditions are satisfied for almost all parameter values; for practical simulation purposes, they can be enforced by arbitrarily small perturbations of the parameter $\alpha$ that do not affect the qualitative dynamics.

The following theorem requires these conditions to hold for some integers $k \geq 2$.
It is therefore sufficient—and in practice straightforward—to verify that they hold for a given system for the cases $k = 1,2$.

For the remainder of this section, assume that \eqref{bifode2} satisfies the \emph{$k$-Sternberg nonresonance condition} for some integer $k\geq 2$. Then the following theorem, first proved in \cite[Section 4]{Takens1971}, holds.
Below we use the notation $C^m$, $m\in \{1,2,\dots\}$, to denote the set of functions that have an $m$-th derivative that is continuous in its domain.  

\begin{theorem}[Takens]\label{thm:takens}
Suppose $\bF$ in \eqref{bifode2} is $C^\infty$ and the above $k$-Sternberg nonresonance condition holds for some $k\geq 2$.
Then there exists a local $C^k$ change of coordinates
\[
(\bx,\alpha)\mapsto (\bu,\bv,\bw,\alpha),
\qquad
\bu\in\bbR^{n_0},\ \bv\in\bbR^{n_-},\ \bw\in\bbR^{n_+},
\]
defined near $(\bx_{\alpha_p},\alpha_p)$, such that the centre manifold $\mathcal{W}^c_\alpha$ in Theorem \ref{thm:man} is mapped precisely onto the linear subspace
\[
\{(\bu,\bv,\bw,\alpha): \ \bv=\bw=0\},
\]
and the extended system \eqref{bifode2} takes the standard form
\begin{equation}\label{eq:takens_standard}
\left\{
\begin{aligned}
\dot \bu &= \bm f(\bu,\alpha),\\
\dot \bv &= \bA_{\text{s}}(\bu,\alpha)\bv,\\
\dot \bw &= \bA_{\text{u}}(\bu,\alpha)\bw,\\
\dot\alpha &= 0,
\end{aligned}
\right.
\end{equation}
where $\bm f,\bA_{\text{s}},\bA_{\text{u}}$ are $C^k$ functions, all eigenvalues of the $n_-\times n_-$ matrix $\bA_{\text{s}}(\bu(t),\alpha)$ have negative real part, and all eigenvalues of the $n_+\times n_+$ matrix $\bA_{\text{u}}(\bu(t),\alpha)$ have positive real part. 

In particular, the reduced dynamics on $\mathcal{W}^c_\alpha$ is represented solely by the evolution of the centre coordinates $\bu$, while the stable and unstable coordinates $(\bv,\bw)$ evolve according to the linear equations determined by $\bA_{\text{s}}(\bu,\alpha)$ and $\bA_{\text{u}}(\bu,\alpha)$.
\end{theorem}

\paragraph{Relation to local topological equivalence}
From Theorem~\ref{thm:takens}, there exist neighbourhoods
\[
\mathcal{E}\subset\bbR^n \quad \text{of } \bx_{\alpha_p},
\qquad
\mathcal{F}\subset\bbR^{n_0+n_-+n_+} \quad \text{of the origin},
\]
and an interval $\mathcal{I}\subset\bbR$ containing $\alpha_p$, such that the coordinate transformation
\[
(\bx,\alpha)\longmapsto (\bu,\bv,\bw,\alpha)
\]
is a $C^k$ diffeomorphism from $\mathcal{E}\times\mathcal{I}$ onto $\mathcal{F}\times\mathcal{I}$.

For each fixed $\alpha\in\mathcal{I}$, restricting this transformation to the invariant hyperplane $\Pi_\alpha$
in Theorem \ref{thm:man}, defines a $C^k$ diffeomorphism
\[
\bh_\alpha:\ \mathcal{E}_\alpha \longrightarrow \mathcal{F}_\alpha,
\qquad
\bx \longmapsto (\bu,\bv,\bw),
\]
where 
\[ \mathcal{E}_\alpha:=\mathcal{E}\cap\{\bx:\ (\bx,\alpha)\in\Pi_\alpha\}, \quad \text{and} \quad \mathcal{F}_\alpha:= \bh_\alpha(\mathcal{E}_\alpha). \]
Under this mapping, trajectories of the original system \eqref{eq:bifode1} in $\mathcal{E}_\alpha$ correspond to trajectories of the standard-form system \eqref{eq:takens_standard} in $\mathcal{F}_\alpha$ with the same time orientation. Consequently, for each $\alpha\in\mathcal{I}$, the map $\bh_\alpha$ realises the local topological equivalence of Definition~\ref{def:topequiv} between the original system \eqref{eq:bifode1} in $\mathcal{E}_\alpha$ and the standard-form system \eqref{eq:takens_standard} in $\mathcal{F}_\alpha$.

\paragraph{Eigenspace projection approximation of the centre coordinates}
Although the mapping $\bh_\alpha$ from the original coordinates $\bx$ to the
centre--stable--unstable coordinates $(\bu,\bv,\bw)$ is in general nonlinear, its action
on the centre coordinates admits a simple first--order approximation near the
non--hyperbolic equilibrium $\bx_{\alpha_p}$.
By construction, the centre manifold $\mathcal W^c_\alpha$ is tangent at
$\bx_{\alpha_p}$ to the eigenspace $E^c$ associated with the eigenvalues of
$\bJ(\bx_{\alpha_p},\alpha_p)$ having zero real part.
Consequently, the differential $\nabla_\bx \bh_\alpha(\bx_{\alpha_p})$ restricted to the
centre coordinates coincides with the linear projection onto $E^c$.

Let $\bR_c\in\mathbb R^{n\times n_0}$ denote a matrix whose columns form a basis of this
eigenspace, and define the associated coordinate projection $\bm P_c := ({\bR_c}^\top\bR_c)^{-1}{\bR_c}^\top$ into this basis.
Then, for any state $\bX$ sufficiently close to $\bx_{\alpha_p}$, the centre component
$\bU$ of $\bh_\alpha(\bX)$ satisfies
\begin{equation}
\bU
= \bm P_c(\bX-\bx_{\alpha_p}) + O\left(\|\bX-\bx_{\alpha_p}\|^2\right).
\end{equation}
This approximation shows that, to leading order, the nonlinear centre coordinates are
obtained simply by projecting deviations from the equilibrium onto $E^c$.

\paragraph{Application of It\^o's Lemma to the image of the LNA under the local homeomorphism}
To simplify notation as before, in this section we use the notation $x_t = x(t)$, for each state variable considered (i.e.\ $\bx$, $\bxi$, $\bX$). Under the non-resonance conditions on the eigenvalues of
$\bJ(\bx_{\alpha_p},\alpha_p)$ required for Theorem~\ref{thm:takens}, the mapping $\bh_\alpha:\mathcal{E}_\alpha\to\mathcal{F}_\alpha$ is
$C^k$ for some $k\geq 2$. Given that $\bX_0=\bx_0 + \Omega^{-1/2}\bxi_0\in\mathcal{E}_\alpha$, this level of regularity is sufficient to apply It\^o's Lemma (see \cite[Theorem 5.10]{LeGallsdebook}) to the image of the continuous semimartingale $\bX_t=\bx_t + \Omega^{-1/2}\bxi_t$ under $\bh_\alpha$ and obtain an SDE representation of its evolution for times $t\in[0,\tau)$ where,
\[  
\tau := \inf\{t\geq 0 : \bX_t\notin\mathcal{E}_\alpha\}.
\]
To this end,
\begin{align}\label{eq:ito1}
    d\bh_\alpha(\bX_t) &= \nabla_\bx \bh_\alpha(\bX_t) \ d\bX_t +\frac{1}{2\Omega}\text{Tr}\left(\bE_t^\top\nabla^2_\bx\bh_\alpha(\bX_t)\bE_t\right) \ dt \\
    &= \nabla_\bx \bh_\alpha(\bX_t)\left( \bF(\bx_t,\alpha) + \frac{1}{\sqrt{\Omega}}\bJ_t\bxi_t\right) \ dt \notag \\ 
    &\qquad + \frac{1}{\sqrt{\Omega}}\nabla_\bx \bh_\alpha(\bX_t)\bE_t \ d\bB_t + \frac{1}{2\Omega}\text{Tr}\left(\bE_t^\top\nabla^2_\bx\bh_\alpha(\bX_t)\bE_t\right) \ dt. \notag
\end{align}
Taylor expanding $\nabla_\bx \bh_\alpha(\bX_t)=\nabla_\bx \bh_\alpha(\bx_t+\Omega^{-1/2}\bxi_t)$ about $\bx_t$ gives
\[ \nabla_\bx \bh_\alpha(\bX_t) = \nabla_\bx \bh_\alpha(\bx_t) + \frac{1}{\sqrt{\Omega}}\bxi_t^\top\nabla^2_\bx\bh_\alpha(\bx_t) + O(\Omega^{-1}), \]
and so \eqref{eq:ito1} becomes,
\begin{align}\label{eq:ito2}
    d\bh_\alpha(\bX_t) &= \Bigg( \nabla_\bx \bh_\alpha(\bx_t)\bF(\bx_t,\alpha) + \frac{1}{\sqrt{\Omega}}\bxi^\top_t\nabla^2_\bx\bh_\alpha(\bx_t)\bF(\bx_t,\alpha) \\
    &\qquad+ \frac{1}{\sqrt{\Omega}}\nabla_\bx \bh_\alpha(\bx_t)\bJ_t\bxi_t\Bigg) \ dt + \frac{1}{\sqrt{\Omega}}\nabla_\bx \bh_\alpha(\bx_t)\bE_t \ d\bB_t + O(\Omega^{-1}). \notag
\end{align}
By Theorem \ref{thm:takens}, $(\bu_t,\bv_t,\bw_t)^\top:=\bh_\alpha(\bx_t)$ evolves according to \eqref{eq:takens_standard} and hence 
\begin{equation}\label{eq:itosub1_1}
\frac{d}{dt}(\bu_t,\bv_t,\bw_t) = \frac{d}{dt} \bh_\alpha(\bx_t) =\nabla_\bx\bh_\alpha(\bx_t) \frac{d}{dt}\bx_t = \nabla_\bx\bh_\alpha(\bx_t) \bF(\bx_t,\alpha), 
    % \nabla_\bx\bh_\alpha(\bx_t)\bF(\bx_t,\alpha) = \frac{d}{dt}\bh_\alpha(\bx_t) = \begin{pmatrix}
    % \bm f(\bu_t,\alpha) \\
    % \bA_{\text{s}}(\bu_t,\alpha)\bv_t \\ 
    % \bA_{\text{u}}(\bu_t,\alpha)\bw_t
% \end{pmatrix} = \widehat \bF (\bh_\alpha(\bx_t))
\end{equation}
which enables to write the first term in the RHS of \eqref{eq:ito2}
% we may write
\begin{equation}\label{eq:itosub1}
\nabla_\bx\bh_\alpha(\bx_t)\bF(\bx_t,\alpha) = \frac{d}{dt}\bh_\alpha(\bx_t) = 
\begin{pmatrix}
    \bm f(\bu_t,\alpha) \\
    \bA_{\text{s}}(\bu_t,\alpha)\bv_t \\ 
    \bA_{\text{u}}(\bu_t,\alpha)\bw_t
\end{pmatrix} = \widetilde \bF_\alpha (\bh_\alpha(\bx_t))
\end{equation}

where
\begin{equation}
    \widetilde \bF_\alpha:\ \begin{pmatrix}
    \bu \\ \bv \\ \bw \end{pmatrix} \longmapsto \begin{pmatrix}
    \bm f(\bu,\alpha) \\
    \bA_{\text{s}}(\bu,\alpha)\bv \\ 
    \bA_{\text{u}}(\bu,\alpha)\bw
\end{pmatrix}.
\end{equation}
Then, using the product rule of differentiation, the second and third term in \eqref{eq:ito2} are
\begin{align}\label{eq:itosub2}
\bxi^\top_t\nabla^2_\bx\bh_\alpha(\bx_t)\bF(\bx_t,\alpha) &+ \nabla_\bx \bh_\alpha(\bx_t)\bJ_t\bxi_t  \notag \\
    &\qquad = \nabla_\bx \left(\nabla_\bx\bh_\alpha(\bx_t)\bF(\bx_t,\alpha)\right)\bxi_t \notag \\
     &\qquad= \nabla_{(\bu,\bv,\bw)}\left( \widetilde \bF_\alpha(\bh_\alpha(\bx_t)) \right)\bxi_t \notag \\
     &\qquad= \begin{pmatrix}
        \nabla_\bu \bm f(\bu_t,\alpha) & 0 & 0 \\
        \nabla_\bu(\bA_{\text{s}}(\bu_t,\alpha)\bv_t) & \bA_{\text{s}}(\bu_t,\alpha) & 0 \\
        \nabla_\bu (\bA_{\text{u}}(\bu_t,\alpha)\bw_t) & 0 &  \bA_{\text{u}}(\bu_t,\alpha)
    \end{pmatrix} \nabla_\bx\bh_\alpha(\bx_t)\bxi_t  \notag \\
     &\qquad= \begin{pmatrix}
        \nabla_\bu \bm f(\bu_t,\alpha) \bm \xi_t^{\bu} \\
        \nabla_\bu(\bA_{\text{s}}(\bu_t,\alpha)\bv_t)\bm \xi_t^{\bu} + \bA_{\text{s}}(\bu_t,\alpha)\bm \xi_t^{\bv} \\
        \nabla_\bu(\bA_{\text{u}}(\bu_t,\alpha)\bw_t)\bm \xi_t^{\bu} + \bA_{\text{u}}(\bu_t,\alpha)\bm \xi_t^{\bw}
    \end{pmatrix}
\end{align}
where $(\bm \xi^{\bu}_t,\bm \xi^{\bv}_t,\bm \xi^{\bw}_t)^\top:=\nabla_\bx\bh_\alpha(\bx_t)\bxi_t$. Finally, by
substituting \eqref{eq:itosub1} and \eqref{eq:itosub2} into \eqref{eq:ito2} we have,
\begin{multline}\label{eq:ito3}
    d\bh_\alpha(\bX_t) = \begin{pmatrix}
    \bm f(\bu_t,\alpha) \\
    \bA_{\text{s}}(\bu_t,\alpha)\bv_t \\ 
    \bA_{\text{u}}(\bu_t,\alpha)\bw_t
    \end{pmatrix} \ dt + \frac{1}{\sqrt{\Omega}}\begin{pmatrix}
        \nabla_\bu \bm f(\bu_t,\alpha) \bm \xi_t^{\bu} \\
        \nabla_\bu(\bA_{\text{s}}(\bu_t,\alpha)\bv_t)\bm \xi_t^{\bu} + \bA_{\text{s}}(\bu_t,\alpha)\bm \xi_t^{\bv} \\
        \nabla_\bu(\bA_{\text{u}}(\bu_t,\alpha)\bw_t)\bm \xi_t^{\bu} + \bA_{\text{u}}(\bu_t,\alpha)\bm \xi_t^{\bw}
    \end{pmatrix} \ dt \\
    + \frac{1}{\sqrt{\Omega}}\nabla_\bx \bh_\alpha(\bX_t)\bE_t \ d\bB_t + O(\Omega^{-1}).
\end{multline}
From inspecting the form of \eqref{eq:ito3}, we conclude that
\begin{equation}
    \bh_\alpha(\bX_t) = \begin{pmatrix}
        \bu_t \\ \bv_t \\ \bw_t
    \end{pmatrix} + \frac{1}{\sqrt{\Omega}}\begin{pmatrix}
        \bm \xi_t^{\bu} \\ \bm \xi_t^{\bv} \\ \bm \xi_t^{\bw}
    \end{pmatrix} + O(\Omega^{-1}),
\end{equation}
where $\bu_t,\bv_t,\bw_t$ evolve according to \eqref{eq:takens_standard} and $\bm \xi^{\bu}_t,\bm \xi^{\bv}_t,\bm \xi^{\bw}_t$ evolve according to the following system of linear SDEs
\begin{equation}
    \left\{
    \begin{aligned}
        d \bm \xi^{\bu}_t &= \nabla_\bu \bm f(\bu_t,\alpha)\bm \xi^{\bu}_t \ dt + \bH^\bu_t \ d\bB_t,\\
        d \bm \xi^{\bv}_t &= \left(\nabla_\bu (\bA_{\text{s}}(\bu_t,\alpha)\bv_t)\bm \xi^{\bu}_t + \bA_{\text{s}}(\bu_t,\alpha) \bm \xi^{\bv}_t\right) \ dt + \bH^\bv_t \ d\bB_t,\\
        d \bm \xi^{\bw}_t &= \left(\nabla_\bu (\bA_{\text{u}}(\bu_t,\alpha)\bw_t)\bm \xi^\bu_t + \bA_{\text{u}}(\bu_t,\alpha) \bm \xi^{\bw}_t\right) \ dt + \bH^{\bw}_t \ d\bB_t,
    \end{aligned}
    \right.
\end{equation}
with 
\begin{equation}
    \begin{bmatrix}
        \bH^\bu_t \\ \bH^\bv_t \\ \bH^{\bw}_t
    \end{bmatrix}
    := \nabla_\bx \bh_\alpha(\bx_t)\bE_t.
\end{equation}
  
\section{Numerical Investigations} \label{SIsec:numerics}

In this section, we give further details related to the numerical investigations included in Section \ref{sec:numerical_investig}. 
In particular, we provide the details of the systems used and additional comparisons between SSA and pcLNA. 

\subsection{The NF-$\kappa$B reaction network}\label{sisec:nfkb}

The NF-$\kappa$B network consists of $11$ species detailed in Table \ref{sitab:NFkbSpecies}.

\begin{table}[H]
\centering
\caption{\label{sitab:NFkbSpecies} Details of the $11$ species characterising the NF-$\kappa$B network and their initial concentrations in three simulation settings with different value of the bifurcation parameter.}
    \begin{tabular}{ |c|c|c|c|c|c|  }
    
    \hline
     
   & Name   & Description &  Initial values  &  Initial values   &  Initial values  \\
   & &  & Fig.\ \ref{sifig:nfkb015} & Fig.\ \ref{sifig:nfkbBif} & Fig.\  \ref{sifig:nfkb1}\\
        % \hhline{|=|=|=|=|}
        \hline
 1 & $N_c$ & Free Cytoplasmic NFkB & 0.0059 & 0.0065 & 0.0044\\
2 & $I_c$ & Free Cytoplasmic IkBa & 0.0119 & 0.0289 & 0.0267\\
3 & $NI_c$ & Cytoplasmic NFkB-IkBa & 0.0096 & 0.0611 & 0.0442\\
4 & $N_n$ & Free nuclear NFkB & 0.0639 & 0.0111 & 0.0305\\
5 & $I_n$ & Free nuclear IkBa & 0.0001 & 0.0010 & 0.0003\\
6 & $NI_n$ & Nuclear NFkB-IkBa & 0.0002 & 0.0005 & 0.0005 \\
7 & $I_m$ & IkBa transcription & 0.0001& 0.0002 & 0.0002\\
8 & $K_n$ & Kinase IKKn & 0.0368 & 0.0071 &0.0015\\
9 & $K_a$ & Kinase IKKa & 0.0083& 0.0034 & 0.0023 \\
10 & $A_m$ & A20 transcription & 0.0001 & 0.0001 & 0.0001 \\
11 & $A$ & A20 & 0.0117 & 0.0139 & 0.0135\\
    \hline 
    \end{tabular}
\end{table}
    
\begin{table}[H]                                  
\centering    
\caption{The parameters of NF-$\kappa$B system and the values used in simulations. Each of the three different levels of TNF$\alpha$ displayed in last row are used for the simulations in Figures \ref{sifig:nfkb015},\ref{sifig:nfkbBif},\ref{sifig:nfkb1}, respectively.}\label{sitab:NFKBparams}

\begin{tabular}{|l|c|r|l|}                                
\hline
parameter & description & value & unit \\ \hline
$k_v$&  Cytoplasm:Nucleus ratio  & 3.3 & --\\ %\hline
$k_p$&  IKKn production &0.0006 &$s^{-1}$\\ %\hline
$k_a$&  Activation caused by TNFa &0.004 &$s^{-1}$\\ %\hline
$k_i$&  Spontaneous IKK activation &0.003 &$s^{-1}$\\ %\hline
$k_{a1a}$&    NFkB-IkBa association  & 0.5 & $\mu M^{-1}s^{-1}$ \\ %\hline
$k_{d1a}$&    NFkB-IkBa dissociation &   0.0005 & $s^{-1}$ \\%\hline
$k_{c1a}$&    Catalysis of IKK-IkBa dimer &    0.074 &$s^{-1}$\\%\hline
$k_{c2a}$&    Catalysis of IKK-IkBa-NFkB trimer &    0.37 &$s^{-1}$\\%\hline
$k_{t2a}$&  degradation of IkBa (IKK dependent) &0.1 &$s^{-1}$\\%\hline
$c_{4a}$&   Free IkBa degradation & 0.0005 &$s^{-1}$\\%\hline
$c_{5a}$&   NFkB complexed IkBa degradation & 0.000022 &$s^{-1}$\\%\hline
$k_{i1}$&   NFkB nuclear import & 0.0026 &$s^{-1}$\\%\hline
$k_{e1}$&   NFkB nuclear export & 0.000052 &$s^{-1}$\\%\hline
$k_{e2a}$&  NFkB-IkBa nuclear export &0.01&$s^{-1}$\\%\hline
$k_{i3a}$&  IkBa nuclear import &0.00067&$s^{-1}$\\%\hline
$k_{e3a}$& IkBa nuclear export &0.000335&$s^{-1}$\\%\hline
$h$&   Order of hill function &   2 & --\\%\hline
$k$&   Hill constant &  0.0650 & $\mu$M/L\\%\hline
$c_{1a}$&   IkBa mRNA synthesis &1.400e-07 &$\mu M^{-1}s^{-1}$\\%\hline
$c_{2a}$&      IkBa translation rate &0.5 &$s^{-1}$\\%\hline
$c_{3a}$&      IkBa mRNA degradation &0.0003 &$s^{-1}$\\%\hline
$c_{1}$&    IkBa mRNA synthesis & 1.4e-07 & $\mu M^{-1}s^{-1}$\\%\hline
$c_{2}$&     A20 mRNA translation & 0.5 &$s^{-1}$\\%\hline
$c_{3}$&       A20 mRNA degradation & 0.00048&$s^{-1}$\\%\hline
$c_{4}$&        A20 degradation & 0.0045 &$s^{-1}$\\%\hline
$k_{bA20}$&  Half-max A20 inhibition concentration &0.0018 & $\mu$M/L \\%\hline
TNF-$\kappa$B & total NF-$\kappa$B concentration & 0.08 & $\mu M$\\%\hline
TIKK & total IKK concentration & 0.08 & $\mu M$ \\
TNF$\alpha$ & Tumor necrosis factor alpha level & 1.5 / 3.66 / 10 & ng/mL\\
\hline
\end{tabular}

\end{table}

The ODE system for the NF-$\kappa$B system are given in table \ref{sitab:nfkbODEs}

\begin{table}[H]
\centering
\caption{The RRE ODE system equations for the NF-$\kappa$B system in \cite{ashall} }\label{sitab:nfkbODEs}
\begin{tabular}{|c l|}
\hline
      % \begin{align*}
$\dot{ N_c} $ & $= k_{d1a} NI_c  - k_{a1a} N_c I_c  - k_{i1} N_c + c_{5a} NI_c  + k_vk_{e1} N_n + k_{t2a}\times (TNFKB-N_c-NI_c- N_n + NI_n)$  \\[0.5em] %Free Cytoplasmic NFkB 
$\dot{ I}_c $ & $=    k_{d1a} NI_c  - k_{a1a} N_c I_c  - k_{i3a} I_c  + k_{v}k_{e3a} I_n - c_{4a}I_c  + c_{2a} I_m  - k_{c1a} K_a  I_c $ \\[0.5em]  %Free Cytoplasmic IkBa
$\dot{NI}_c $& $=  k_{a1a} N_cI_c  - k_{d1a} NI_c  + k_vk_{e2a} NI_n - c_{5a} NI_c - k_{c2a} K_a   NI_c $ \\[0.5em]  %Cytoplasmic NFkB-IkBa
$\dot{ N}_n $& $=  k_{d1a} NI_n   - k_vk_{a1a} N_nI_n + k_{i1} N_c  - k_{v} k_{e1} N_n  $ \\[0.5em]  %Free nuclear NFkB
$\dot{ I}_n $& $=  k_{d1a} NI_n   - k_vk_{a1a} N_nI_n      + k_{i3a} I_c  - k_{v} k_{e3a} I_n - c_{4a} I_n     $ \\[0.5em]  %Free nuclear IkBa
\nonumber $\dot{ NI}_n $& $=  k_vk_{a1a} NI_n - k_{d1a} NI_n  - k_{v} k_{e2a} NI_n  $ \\[0.5em]  %Nuclear NFkB-IkBa
$\dot{ I}_m $& $=  c_{1a} (N_n  ^h/(N_n  ^h + (k/k_v)^h)) - c_{3a} I_m $ \\[0.5em]  %IkBa transcription
$\dot{ K_n} $& $=  k_{p} (TIKK - K_n - K_a)    (k_{bA20}/(k_{bA20} + A \times TNF\alpha/10)) - k_{a} TNF\alpha  K_n/10 $ \\[0.5em]  %K_n
$\dot{ K}_a $& $=  k_{a} TNF\alpha K_n/10  - k_{i} K_a  $ \\[0.5em]  %K_a
$\dot{ A}_m $& $=  c_{1} (N_n  ^h/(N_n  ^h + (k/k_v)^h)) - c_{3} A_m $ \\[0.5em]  %A transcription
$\dot{ A} $& $=  c_{2} A_m  - c_{4} A $ \\[0.5em]  %A
% \end{align*}
\hline
\end{tabular}
\end{table}

The propensity functions used for the SSA are provided in Table \ref{sitab:NFKBrates}. 
The values of the parameters are the same as in Table \ref{sitab:NFKBparams} with suitable rate constant conversions applied.\\
\begin{table}[H]\label{sitab:NFKBrates}                   \centering           
\caption{NF-$\kappa$B reactions and the corresponding propensity functions used for the SSA simulations of the NF-$\kappa$B system.}\label{sitab:nfkbrates}
\begin{tabular}{|c|c|}\hline
\centering
reaction & rate \\ \hline
$N_c + I_c \xrightarrow{k_{a1a}} NI_c$ & $k_{a1a} \times I_c \times N_c / \Omega$\\\hline
$NI_c \xrightarrow{k_{d1a}} N_c + I_c$ & $k_{d1a} \times NI_c$ \\\hline
$N_n + I_n \xrightarrow{k_{a1a}} NI_n$ & $k_vk_{a1a} \times I_n \times N_n / \Omega$ \\ \hline
$NI_n \xrightarrow{k_{d1a}} N_n+I_n $ & $k_{d1an} \times NI_n$ \\\hline
$K_a + I_c \xrightarrow{k_{c1a}} K_a + I_p $ & $k_{c1a} \times K_a \times I_c / \Omega$ \\\hline
$K_a + NI_c \xrightarrow{k_{c2a}} K_a + NI_p $ & $k_{c2a} \times K_a \times NI_c / \Omega$  \\\hline
$NI_p \xrightarrow{k_{t2a}} N_c$ & $k_{t2a} \times NI_p$ \\\hline
$N_c \xrightarrow{k_{i1}} N_n $ & $ k_{i1} \times N_c$  \\\hline
$N_n \xrightarrow{k_{e1}} N_c $ & $ k_{e1} \times k_v \times N_n$  \\\hline
$NI_n  \xrightarrow{k_{e2a}} NI_c $ & $k_{e2a} \times k_v \times NI_n$ \\\hline
$I_c \xrightarrow{k_{i3a}}  I_n$ & $k_{i3a} \times I_c$  \\\hline
$I_n \xrightarrow{k_{e3a}} I_c $ & $k_{e3a} \times k_v \times I_c$  \\\hline
$\varnothing \xrightarrow{H_I} I_m $ & $ (c_{1a} \Omega)\frac{N_n^h}{N_n^h+(k \Omega/k_v)^h} $ \\\hline
$I_m \xrightarrow{c_{2a}} I_m + I_c$ & $c_{2a} \times I_m$ \\\hline
$I_m \xrightarrow{c_{3a}} \varnothing $ & $c_{3a} \times I_m$ \\\hline
$I_c \xrightarrow{c_{4a}} \varnothing$ & $c_{4a} \times I_c$ \\\hline
$I_n \xrightarrow{c_{4a}} \varnothing$ & $c_{4a} \times I_n$ \\\hline
$NI_c \xrightarrow{c_{5a}} N_c$ & $c_{5a} \times NI_c$\\\hline
$\varnothing \xrightarrow{H_A} A_m $ & $ (c_{1} \Omega) \frac{N_n^h}{N_n^h+(k \Omega/k_v)^h}$ \\\hline
$A_m \xrightarrow{c_2} A_m +A$ & $c_2 \times A_m$ \\\hline
$A_m \xrightarrow{c_3} \varnothing$ & $c_3 \times A_m$ \\\hline
$A \xrightarrow{c_4} \varnothing$ & $c_4 \times A$ \\\hline
$K_i \xrightarrow{M_A} K_n $ & $k_p  (TIKK - K_n - K_a) \frac{k_{bA20} \times \Omega }{(k_{bA20} \Omega)+ A \times TNF\alpha/10} $ \\\hline
$K_n \xrightarrow{TNF\alpha/10 \times k_a} K_a $ & $ TNF\alpha/10 \times k_a \times K_n $\\\hline
$K_a \xrightarrow{k_i} K_i $ & $k_i \times K_a$ \\
\hline
\end{tabular}
\end{table}

\subsubsection{Details for Figure \ref{fig:Numerical}}\label{SIsec:Fig3A}

The bifurcation parameter of this system is the level of the Tumor necrosis factor alpha level (TNF$\alpha$). 
The Hopf bifurcation value is $\approx 3.66 ng/mL$. 
The simulations in Figure \ref{fig:Numerical} are produced with the level of TNF$\alpha$ at the Hopf bifurcation value and with initial conditions given in Table \ref{sitab:NFkbSpecies} in the column ``Initial values  Fig.\ \ref{sifig:nfkbBif}''. 

The empirical density functions of the probability distribution of each variable are derived using the kernel density function ``ksdensity'' in MATLAB \cite{MATLAB}. 
To ensure good visibility of all the curves, we scale them. 
Specifically, both the $x$- and $z$-axis range from $0$ to $1$ since we divide both the value of each variable and their densities obtained for both SSA and pcLNA by their maximum observed values for both algorithms. 

\subsubsection{Further numerical investigations for the NF-$\kappa$B system} \label{SIsec:nfkbNumerical}

\begin{figure}[H]
\centering
\includegraphics[width=\textwidth]{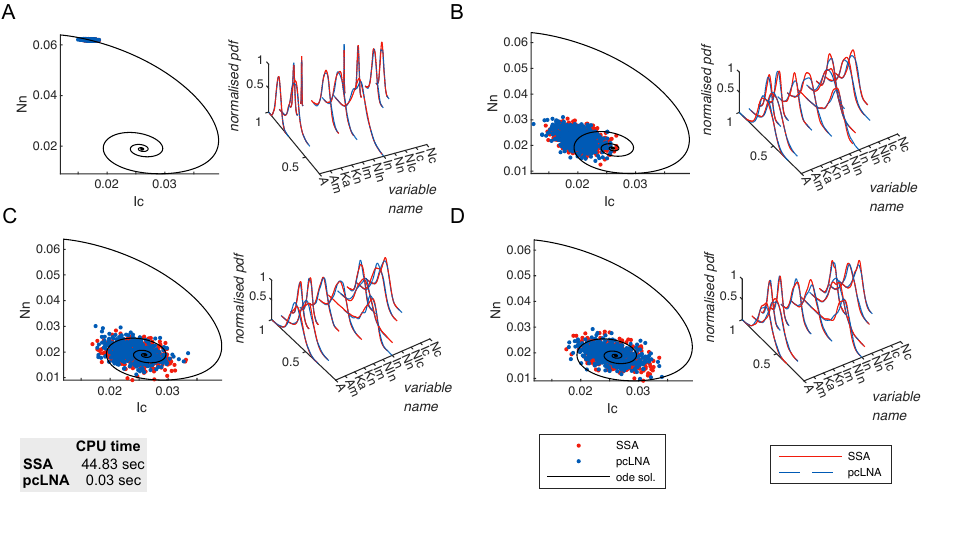}
\caption{\label{sifig:nfkb015} 
Comparison between SSA (red) and pcLNA (blue) for the NF-$\kappa$B system at $\Omega = 600{,}000$ with \TNFA level $= 1.5$ ng/mL. 
Panels (A)–(D) correspond to times $t=5,\ \tau+5,\ 2\tau+5,\ 3\tau+5$, where $\tau \approx 102.095$ is the oscillation period. 
Each panel shows (left) the states recorded at time $t$ of the simulations with solutions of the corresponding RRE, and (right) the corresponding empirical probability density functions of each system variable. 
Initial conditions and parameter values are given in Tables~\ref{sitab:NFkbSpecies} and~\ref{sitab:NFKBparams}.
}
\end{figure}

\begin{figure}[H]
\centering
\includegraphics[scale=0.9]{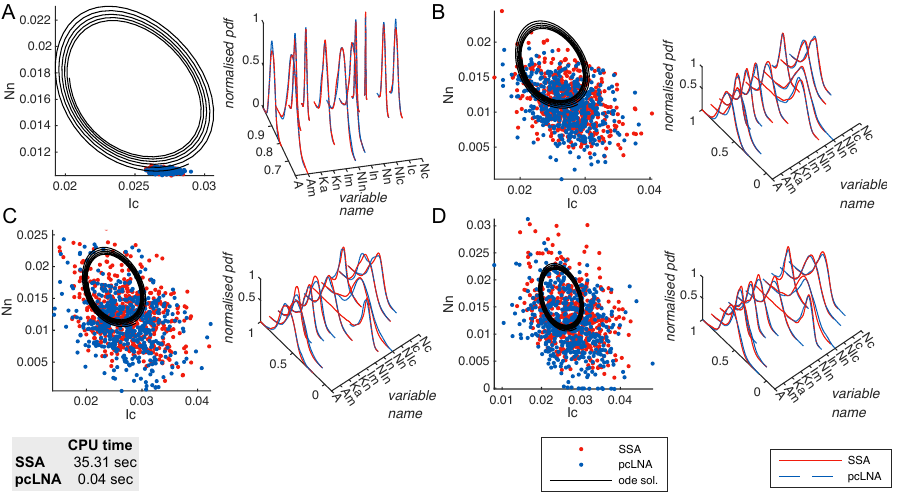}
\caption{\label{sifig:nfkbBif} 
Comparison between SSA (red) and pcLNA (blue) for the NF-$\kappa$B system at $\Omega = 600{,}000$ with \TNFA level $= 3.66$ ng/mL (Hopf bifurcation point). 
Panels (A)–(D) correspond to times $t=5,\ \tau+5,\ 2\tau+5,\ 3\tau+5$, where $\tau \approx 102.095$ is the oscillation period. 
Each panel shows (left) the states recorded at time $t$ of the simulations with solutions of the corresponding RRE, and (right) the corresponding empirical probability density functions of each system variable. 
Initial conditions and parameter values are provided in Tables~\ref{sitab:NFkbSpecies} and~\ref{sitab:NFKBparams}.
}
\end{figure}

\begin{figure}[H]
\centering
\includegraphics[scale=0.9]{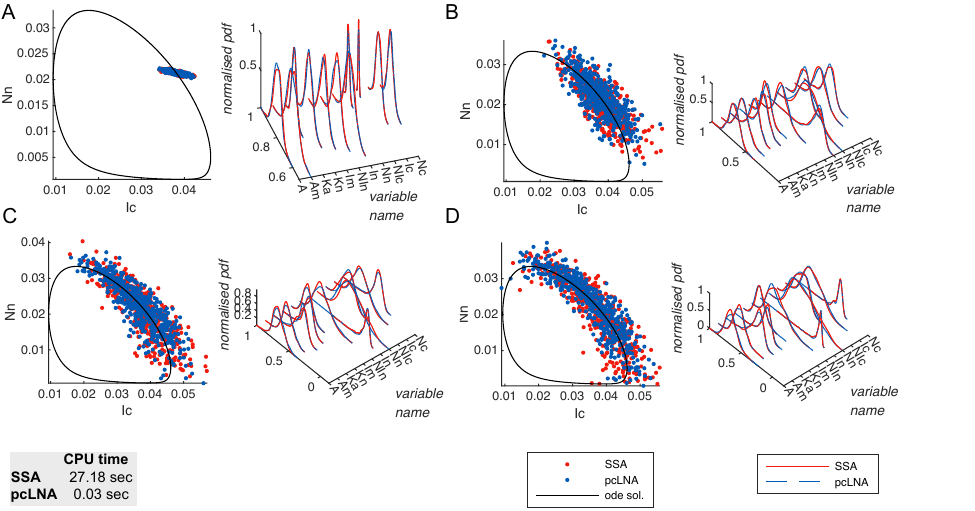}
\caption{\label{sifig:nfkb1} 
Comparison between SSA (red) and pcLNA (blue) for the NF-$\kappa$B system at $\Omega = 600{,}000$ with \TNFA level $= 10$ ng/mL. 
Panels (A)–(D) correspond to times $t=5,\ \tau+5,\ 2\tau+5,\ 3\tau+5$, where $\tau \approx 102.095$ is the oscillation period. 
Each panel shows (left) the states recorded at time $t$ of the simulations with solutions of the corresponding RRE, and (right) the corresponding empirical probability density functions of each system variable. 
Initial conditions and parameter values are listed in Tables~\ref{sitab:NFkbSpecies} and~\ref{sitab:NFKBparams}.
}
\end{figure}

See caption of Figures \ref{sifig:nfkb015}, \ref{sifig:nfkbBif}, \ref{sifig:nfkb1} for more details on additional comparisons for this system, in the cases where the parameter values give a system that presents a stable focus (see Figure \ref{sifig:nfkb015}), the case where the parameter values are exactly on the Hopf bifurcation point (see Figure \ref{sifig:nfkbBif}), and the case where a limit cycle appears (see Figure \ref{sifig:nfkb1}).

\subsection{The Brusselator reaction network}\label{sisec:brus}

The two dimensional Brusselator network, introduced in \cite{LEFEVER1971267}, consists of two chemical species, $A$ and $B$, which undergo the following reactions, 
\[ \emptyset \xrightarrow{1} A, \qquad A \xrightarrow{1} \emptyset, \qquad A \xrightarrow{b} B, \qquad 2A+B \xrightarrow{c} 3A.\] 
If, for each $t\geq 0$, we let $\bY(t) = (Y_1(t),Y_2(t))^\top$ denote the number of molecules of $A$ and $B$ respectively, then these reactions occur according to the following intensities 
\[ \pi_1=\Omega, \qquad \pi_2=Y_1, \qquad \pi_3=bY_1, \qquad \pi_4 = cY_1^2Y_2/\Omega^2.\]
With system size $\Omega$ and $\bX(t) = \bY(t)/\Omega$ for each $t\geq 0$, the corresponding RRE solution in the $\Omega \to \infty$ limit is
\begin{equation}\label{brusRRE}
    \left\{
    \begin{aligned}
    \dot{x_1} &= 1 - x_1(1+b-cx_1x_2), \\
    \dot{x_2} &= x_1(b-cx_1x_2).
    \end{aligned}
    \right.
\end{equation}

From studying the Jacobian matrix of \eqref{brusRRE}, we may conclude the system has a non-hyperbolic equilibrium at $(x_1,x_2) = (1,2)^\top$, when $b=2$ and $c=1$, with purely imaginary conjugate eigenvalues $\pm i$. Through small deviations of $b$ we deduce the Hopf bifurcation is supercritical with losses of stability occurring as the eigenvalues cross the imaginary axis. Parameters $b<2$ give rise to a stable equilibrium, whereas $b>2$ gives rise to an unstable equilibrium surrounded by a unique and stable limit cycle. 

% Species Table
\begin{table}[H]
\centering
\caption{\label{tab:brusselatorSpecies} Species in the Brusselator network and their initial concentrations for three different values of parameter \( b \).}
\begin{tabular}{ |c|c|c|c|c|c| }
\hline
 & Name & Description & Initial values  & Initial values & Initial values  \\
 &  &   & Fig.\ \ref{sifig:brus1-7} &  Fig.\ \ref{sifig:brusBif} & Fig.\ \ref{sifig:brus2-3} \\
\hline
1 & $A$ & Chemical species A & 0.8000 & 0.8000 & 1.0340 \\
2 & $B$ & Chemical species B & 1.5000 & 1.5000 & 2.9230 \\
\hline
\end{tabular}
\end{table}

% Parameter Table
\begin{table}[H]
\centering
\caption{\label{tab:brusselatorParams} Parameters used in the Brusselator network.}
\begin{tabular}{|l|c|r|l|}
\hline
Parameter & Description & Value & Unit \\
\hline
$c$ & Catalytic reaction rate & 1 & $\mu M^{-2}s^{-1}$ \\
$b$ & Conversion rate of A to B & 1.7 / 2.0 / 2.3 & $s^{-1}$ \\
\hline
\end{tabular}
\end{table}

% ODE System
\begin{table}[H]
\centering
\caption{\label{tab:brusselatorODEs} The RRE ODEs for the mean-field dynamics of the Brusselator in the limit $\Omega \to \infty$.}
\begin{tabular}{|c l|}
\hline
$\dot{x}_1$ & $= 1 - x_1(1 + b - c x_1 x_2)$ \\[0.5em]
$\dot{x}_2$ & $= x_1(b - c x_1 x_2)$ \\
\hline
\end{tabular}
\end{table}

% SSA Propensity Table
\begin{table}[H]
\centering
\caption{\label{tab:brusselatorSSA} Brusselator reactions and their corresponding propensity functions for SSA simulations.}
\begin{tabular}{|c|l|}
\hline
Reaction & Propensity function \\
\hline
$\varnothing \xrightarrow{1} A$ & $\pi_1 = \Omega$ \\
$A \xrightarrow{1} \varnothing$ & $\pi_2 = Y_1$ \\
$A \xrightarrow{b} B$ & $\pi_3 = b Y_1$ \\
$2A + B \xrightarrow{c} 3A$ & $\pi_4 = c Y_1^2 Y_2/\Omega^2$ \\
\hline
\end{tabular}
\end{table}

\begin{figure}[H]
\centering
\includegraphics[scale=0.9]{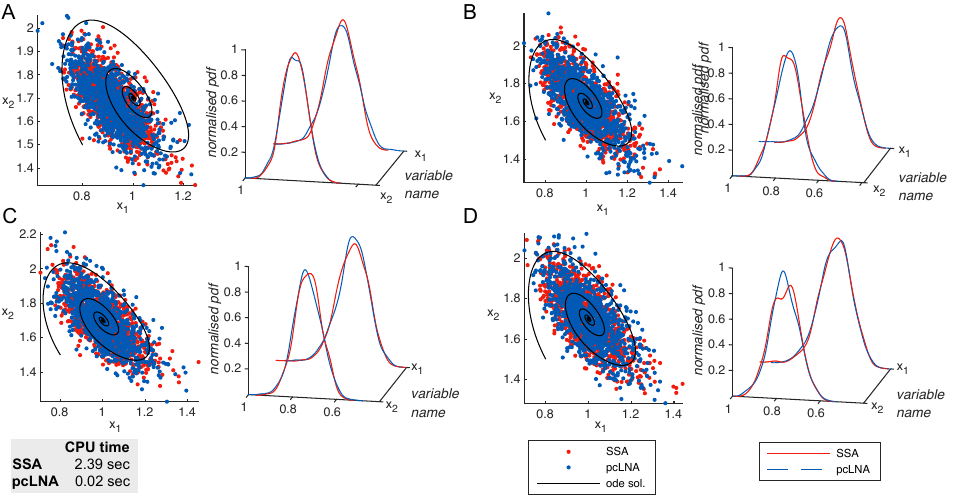}
\caption{\label{sifig:brus1-7} 
Comparison between SSA (red) and pcLNA (blue) for the Brusselator system at $\Omega = 1000$ with parameter $b=1.7$. 
Panels (A)–(D) correspond to times $t=\tau+0.5,\ 3\tau+0.5,\ 5\tau+0.5,\ 7\tau+0.5$ with period $\tau = 6.2919$. 
Each panel shows (left) the states recorded at time $t$ of the simulations with solutions of the corresponding RRE, and (right) the corresponding empirical probability density functions of each system variable. 
Initial conditions and parameter values are given in Tables~\ref{tab:brusselatorSpecies} and~\ref{tab:brusselatorParams}.
}
\end{figure}

\begin{figure}[H]
\centering
\includegraphics[scale=0.9]{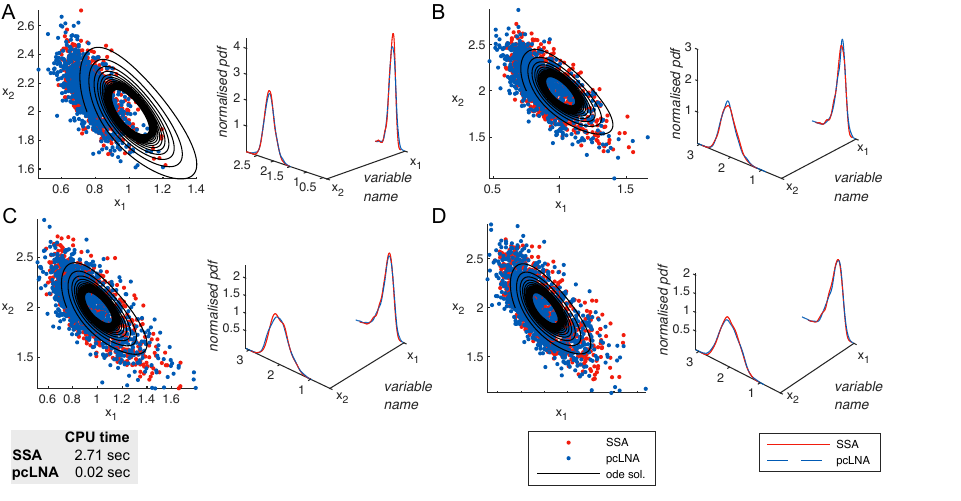}
\caption{\label{sifig:brusBif} 
Comparison between SSA (red) and pcLNA (blue) for the Brusselator system at $\Omega = 1000$ with parameter $b=2$. 
Panels (A)–(D) correspond to times $t=\tau+0.5,\ 3\tau+0.5,\ 5\tau+0.5,\ 7\tau+0.5$ with period $\tau = 6.35$. 
Each panel shows (left) the states recorded at time $t$ of the simulations with solutions of the corresponding RRE, and (right) the corresponding empirical probability density functions of each system variable. 
Initial conditions and parameter values are provided in Tables~\ref{tab:brusselatorSpecies} and~\ref{tab:brusselatorParams}.
}
\end{figure}

\begin{figure}[H]
\centering
\includegraphics[scale=0.9]{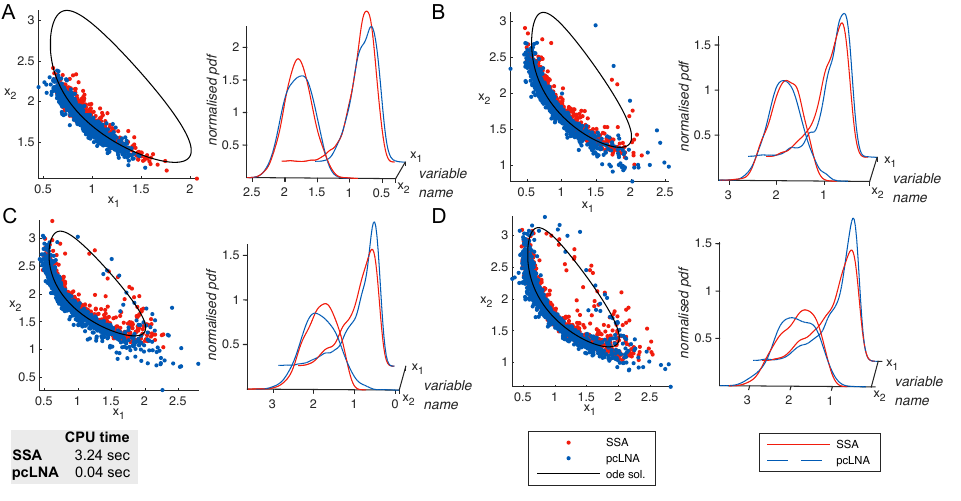}
\caption{\label{sifig:brus2-3} 
Comparison between SSA (red) and pcLNA (blue) for the Brusselator system at $\Omega = 1000$ with parameter $b=2.3$. 
Panels (A)–(D) correspond to times $t=\tau+0.5,\ 3\tau+0.5,\ 5\tau+0.5,\ 7\tau+0.5$ with period $\tau = 6.4276$. 
Each panel shows (left) the states recorded at time $t$ of the simulations with solutions of the corresponding RRE, and (right) the corresponding empirical probability density functions of each system variable. 
Initial conditions and parameter values are listed in Tables~\ref{tab:brusselatorSpecies} and~\ref{tab:brusselatorParams}.
}
\end{figure}

\subsection{The smallest Hopf bifurcation reaction network}\label{sisec:simpleHopf}

\begin{figure}[H]
\centering
\includegraphics[scale=0.9]{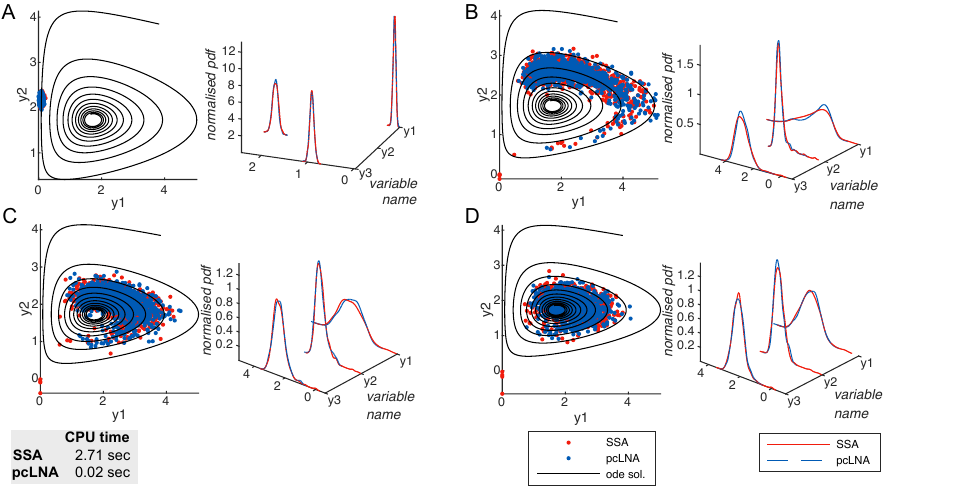}
\caption{\label{sifig:simpleHopf6} 
Comparison between SSA (red) and pcLNA (blue) for the smallest Hopf bifurcation network at $\Omega = 500$ with parameter $k_1 = 6$. 
Panels (A)–(D) correspond to times $t=\tau+1,\ 3\tau+1,\ 5\tau+1,\ 7\tau+1$ with period $\tau \approx 3$. 
Each panel shows (left) the states recorded at time $t$ of the simulations with solutions of the corresponding RRE, and (right) the corresponding empirical probability density functions of each system variable. 
Initial conditions and parameter values are given in Tables~\ref{tab:simpleHopfSpecies} and~\ref{tab:simpleHopfParams}.
}
\end{figure}

\begin{figure}[H]
\centering
\includegraphics[scale=0.9]{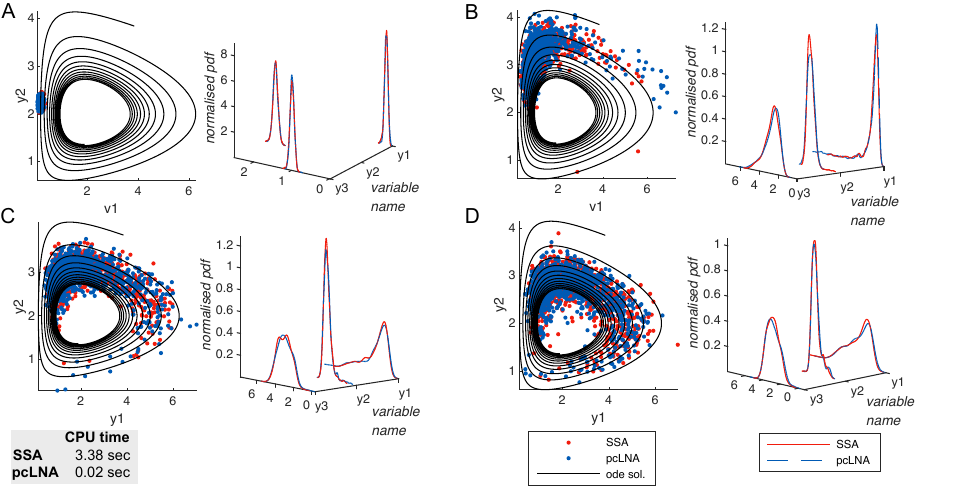}
\caption{\label{sifig:simpleHopfBif} 
Comparison between SSA (red) and pcLNA (blue) for the smallest Hopf bifurcation network at $\Omega = 500$ with parameter $k_1 = 6.6$. 
Panels (A)–(D) correspond to times $t=\tau+1,\ 3\tau+1,\ 5\tau+1,\ 7\tau+1$ with period $\tau \approx 3$. 
Each panel shows (left) the states recorded at time $t$ of the simulations with solutions of the corresponding RRE, and (right) the corresponding empirical probability density functions of each system variable. 
Initial conditions and parameter values are provided in Tables~\ref{tab:simpleHopfSpecies} and~\ref{tab:simpleHopfParams}.
}
\end{figure}

\begin{figure}[H]
\centering
\includegraphics[scale=0.9]{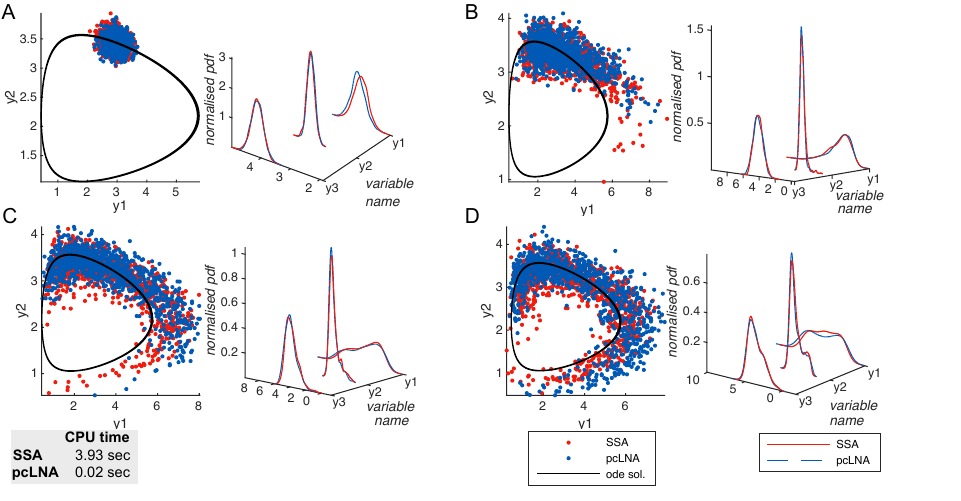}
\caption{\label{sifig:simpleHopf7} 
Comparison between SSA (red) and pcLNA (blue) for the smallest Hopf bifurcation network at $\Omega = 500$ with parameter $k_1 = 7$. 
Panels (A)–(D) correspond to times $t=\tau+1,\ 3\tau+1,\ 5\tau+1,\ 7\tau+1$ with period $\tau \approx 3$. 
Each panel shows (left) the states recorded at time $t$ of the simulations with solutions of the corresponding RRE, and (right) the corresponding empirical probability density functions of each system variable. 
Initial conditions and parameter values are listed in Tables~\ref{tab:simpleHopfSpecies} and~\ref{tab:simpleHopfParams}.
}
\end{figure}

The system is described in \cite{WILHELM1995}. The tables \ref{tab:simpleHopfSpecies}, \ref{tab:simpleHopfParams}, \ref{tab:simpleHopfODEs}, \ref{tab:simpleHopfSSA} describe the species and their initial conditions, the parameters, the ODE system, and the propensity functions of the SSA simulations of this system, respectively. 
The results in Figures \ref{sifig:simpleHopf6}, \ref{sifig:simpleHopfBif}, and \ref{sifig:simpleHopf7} are derived using the same method as used for Figure 3 in the main paper. 
The bifurcation parameter is $k_1$ and its value at bifurcation is $6.6$.
The computation of the empirical density functions is done as described in SM Section~\ref{SIsec:Fig3A}.
We used 1000 trajectories of each simulation algorithm to produce the results. 

% Species Table
\begin{table}[H]
\centering
\caption{\label{tab:simpleHopfSpecies} Species in the smallest Hopf bifurcation network and their initial concentrations for three different values of parameter \( k_1 \).}
\begin{tabular}{ |c|c|c|c|c|c| }
\hline
 & Name & Description & Initial values Fig.\ \ref{sifig:simpleHopf6} & Initial values Fig.\ \ref{sifig:simpleHopfBif} & Initial values Fig.\ \ref{sifig:simpleHopf7} \\
\hline
1 & $X_1$ & Species 1 & 3.8220 & 3.8220 & 2.4460 \\
2 & $X_2$ & Species 2 & 3.8460 & 3.8460 & 1.0720 \\
3 & $X_3$ & Species 3 & 5.1400 & 5.1400 & 1.2940 \\
\hline
\end{tabular}
\end{table}

% Parameter Table
\begin{table}[H]
\centering
\caption{\label{tab:simpleHopfParams} Parameters used in the smallest Hopf bifurcation network.}
\begin{tabular}{|l|c|r|l|}
\hline
Parameter & Description & Value & Unit \\
\hline
$k_1$ & Activation rate & 6 / 6.6 / 7 & $s^{-1}$ \\
$k_2$ & Interaction rate between $X_1$ and $X_2$ & 2.2 & $s^{-1}$ \\
$k_3$ & Degradation rate of $X_2$ & 2.2 & $s^{-1}$ \\
$k_4$ & Conversion rate from $X_1$ to $X_3$ & 2.2 & $s^{-1}$ \\
$k_5$ & Conversion rate from $X_3$ to $X_2$ & 2.2 & $s^{-1}$ \\
$A$ & External activation strength & 1 & -- \\
\hline
\end{tabular}
\end{table}

% ODE System
\begin{table}[H]
\centering
\caption{\label{tab:simpleHopfODEs} The RRE ODEs for the mean-field dynamics of the smallest Hopf bifurcation network in the limit $\Omega \to \infty$.}
\begin{tabular}{|c l|}
\hline
$\dot{x}_1$ & $= k_1 A x_1 - k_4 x_1 - k_2 x_1 x_2$ \\[0.5em]
$\dot{x}_2$ & $= -k_3 x_2 + k_5 x_3$ \\[0.5em]
$\dot{x}_3$ & $= k_4 x_1 - k_5 x_3$ \\
\hline
\end{tabular}
\end{table}

% SSA Propensity Table
\begin{table}[H]
\centering
\caption{\label{tab:simpleHopfSSA} smallest Hopf bifurcation reactions and their corresponding propensity functions for SSA simulations.}
\begin{tabular}{|c|l|}
\hline
Reaction & Propensity function \\
\hline
$X_1 \xrightarrow{k_1 A} X_1$ & $\pi_1 = k_1 A Y_1$ \\
$X_1 + X_2 \xrightarrow{k_2} \emptyset$ & $\pi_2 = k_2 \cdot Y_1 \cdot Y_2 / \Omega$ \\
$X_2 \xrightarrow{k_3} \emptyset$ & $\pi_3 = k_3 \cdot Y_2$ \\
$X_1 \xrightarrow{k_4} X_3$ & $\pi_4 = k_4 \cdot Y_1$ \\
$X_3 \xrightarrow{k_5} X_2$ & $\pi_5 = k_5 \cdot Y_3$ \\
\hline
\end{tabular}
\end{table}

\subsection{The genetic toggle switch reaction network} \label{sisec:toggleswitch}

\begin{figure}[H]
\centering
\includegraphics[scale=0.9]{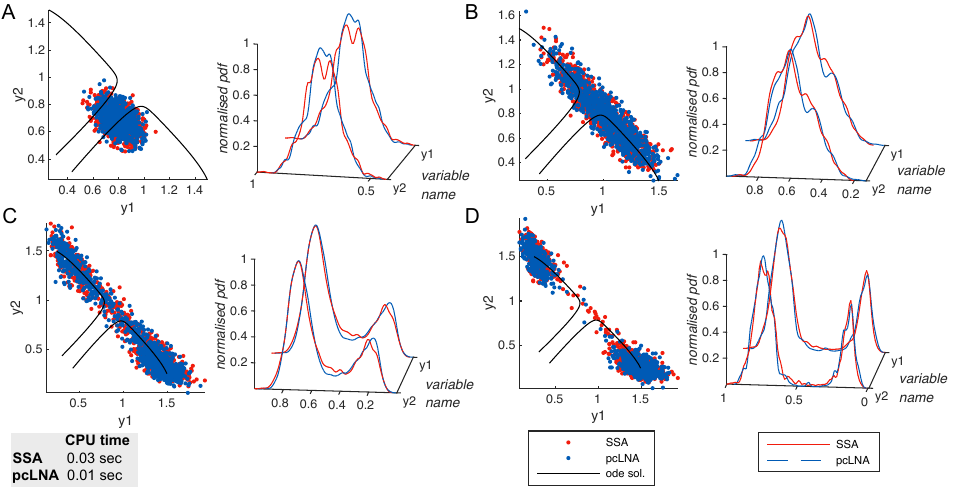}
\caption{\label{sifig:toggleswitch} 
Comparison between SSA (red) and pcLNA (blue) for the genetic toggle switch network at $\Omega = 100$. 
Panels (A)–(D) correspond to times $t=0.5006,\ 2.0024,\ 5.0059,\ 9.5051$, respectively. 
Each panel shows (left) the states recorded at time $t$ of the simulations with solutions of the corresponding RRE, and (right) the corresponding empirical probability density functions of each system variable. 
The initial conditions for the two ODE solutions are provided in Table~\ref{tab:toggleswitchSpecies}, and the parameter values are given in Table~\ref{tab:toggleswitchParams}.
}
\end{figure}

The system is described in \cite{Gardner2000}. The tables \ref{tab:toggleswitchSpecies}, \ref{tab:toggleswitchParams}, \ref{tab:toggleswitchODEs}, \ref{tab:toggleswitchSSA} describe the species and their initial conditions, the parameters, the ODE system, and the propensity functions of the SSA simulations of this system, respectively. 
The results in Figure \ref{sifig:toggleswitch} are derived using the same method as used for Figure 3 in the main paper. 
The computation of the empirical density functions is done as described in SM section~\ref{SIsec:Fig3A}.
We used 1000 trajectories of each simulation algorithm to produce the results. 

\begin{figure}[H]
\centering
\includegraphics[scale=0.9]{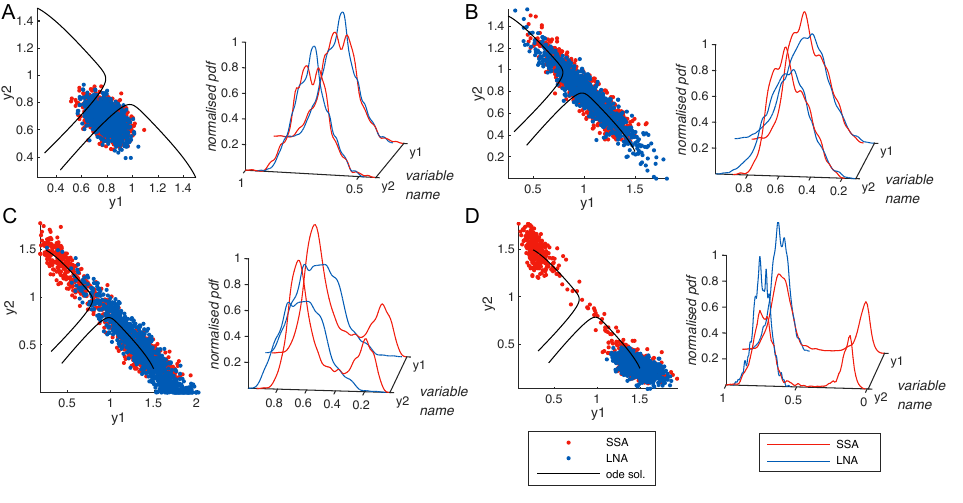}
\caption{\label{sifig:toggleswitchLNA} 
Comparison between SSA (red) and (standard) LNA (blue) for the genetic toggle switch network at $\Omega = 100$. 
Panels (A)–(D) correspond to times $t=0.5006,\ 2.0024,\ 5.0059,\ 9.5051$, respectively. 
Each panel shows (left) the states recorded at time $t$ of the simulations with solutions of the corresponding RRE, and (right) the corresponding empirical probability density functions of each system variable. 
The initial conditions for the two ODE solutions are provided in Table~\ref{tab:toggleswitchSpecies}, and the parameter values are given in Table~\ref{tab:toggleswitchParams}.
}
\end{figure}

The results in Figure \ref{sifig:toggleswitchLNA} are derived using the same method as used for Figure 3 in the main paper, except for the LNA simulation where the steps A and C.3 are omitted (i.e.\ no phase correction is applied). 
The computation of the empirical density functions is done as described in SM section~\ref{SIsec:Fig3A}.
We used 1000 trajectories of each simulation algorithm to produce the results. 

% Species Table
\begin{table}[H]
\centering
\caption{\label{tab:toggleswitchSpecies} Species in the genetic toggle switch network and their initial concentrations.}
\begin{tabular}{ |c|c|c|c|c| }
\hline
 & Name & Description & Initial values  & Initial values \\
 & & & ODE solution 1 Fig.\ \ref{sifig:toggleswitch} &  ODE solution 2 Fig.\ \ref{sifig:toggleswitch} \\
\hline
1 & $x_1$ & Repressor 1 & 0.3067 & 0.4311 \\
2 & $x_2$ & Repressor 2 & 0.4311 & 0.3067 \\
\hline
\end{tabular}
\end{table}

% Parameter Table
\begin{table}[H]
\centering
\caption{\label{tab:toggleswitchParams} Parameters used in the genetic toggle switch network.}
\begin{tabular}{|l|c|r|l|}
\hline
Parameter & Description & Value & Unit \\
\hline
$a_1$ & Activation rate of $x_1$ & 1.5 & $s^{-1}$ \\
$a_2$ & Activation rate of $x_2$ & 1.5 & $s^{-1}$ \\
$b$ & Hill coefficient for $x_2$ regulation of $x_1$ & 4 & -- \\
$g$ & Hill coefficient for $x_1$ regulation of $x_2$ & 4 & -- \\
$A$ & External activation strength & 1 & -- \\
\hline
\end{tabular}
\end{table}

% ODE System
\begin{table}[H]
\centering
\caption{\label{tab:toggleswitchODEs} The RRE ODEs for the mean-field dynamics of the genetic toggle switch in the limit $\Omega \to \infty$.}
\begin{tabular}{|c l|}
\hline
$\dot{x}_1$ & $= \frac{a_1 A}{1 + x_2^b} - x_1$ \\[0.5em]
$\dot{x}_2$ & $= \frac{a_2}{1 + x_1^g} - x_2$ \\
\hline
\end{tabular}
\end{table}

% SSA Propensity Table
\begin{table}[H]
\centering
\caption{\label{tab:toggleswitchSSA} genetic toggle switch reactions and their corresponding propensity functions for SSA simulations.}
\begin{tabular}{|c|l|}
\hline
Reaction & Propensity function \\
\hline
$X_1 \xrightarrow{a_1 A} X_1$ & $\pi_1 = \frac{a_1 A \cdot \Omega}{1 + Y_2^b}$ \\
$X_1 \xrightarrow{} \emptyset$ & $\pi_2 = Y_1$ \\
$X_2 \xrightarrow{a_2} X_2$ & $\pi_3 = \frac{a_2 \cdot \Omega}{1 + Y_1^g}$ \\
$X_2 \xrightarrow{} \emptyset$ & $\pi_4 = Y_2$ \\
\hline
\end{tabular}
\end{table}

\subsection{The cell cycle reaction network} \label{sisec:cellCycle}

\begin{figure}[H]
\centering
\includegraphics[scale=0.9]{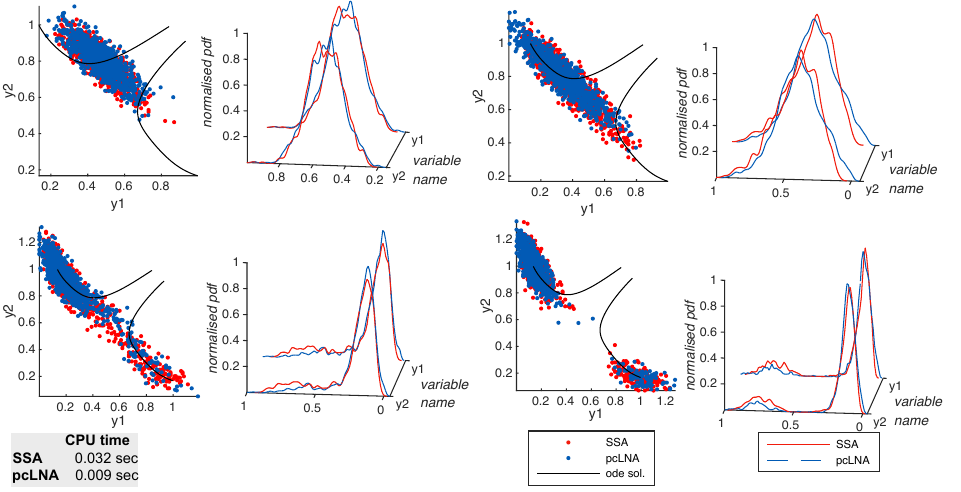}
\caption{\label{sifig:cellcycle} 
Comparison between SSA (red) and pcLNA (blue) for the cell cycle network at $\Omega = 100$. 
Panels (A)–(D) correspond to times $t=0.5000,\ 9.9999,\ 49.9996,\ 100.0000$, respectively. 
Each panel shows (left) the states recorded at time $t$ of the simulations with solutions of the corresponding RRE, and (right) the corresponding empirical probability density functions of each system variable. 
The initial conditions for the two ODE solutions are provided in Table~\ref{tab:cellCycleSpecies}, and the parameter values are given in Table~\ref{tab:cellCycleParams}.
}
\end{figure}

The system is described in \cite{Angeli2004}; see section ``A Two-Variable Example: The Cdc2-Cyclin B/Wee1 System''. 
The tables \ref{tab:cellCycleSpecies}, \ref{tab:cellCycleParams}, \ref{tab:cellCycleODEs}, \ref{tab:cellCycleSSA} describe the species and their initial conditions, the parameters, the ODE system, and the propensity functions of the SSA simulations of this system, respectively. 
The results in Figure \ref{sifig:cellcycle} are derived using the same method as used for Figure 3 in the main paper. 
The computation of the empirical density functions is done as described in SM section~\ref{SIsec:Fig3A}.
We used 1000 trajectories of each simulation algorithm to produce the results. 

% Species Table
\begin{table}[H]
\centering
\caption{\label{tab:cellCycleSpecies} Species in the cell cycle network and their initial concentrations.}
\begin{tabular}{ |c|c|c|c|c| }
\hline
 & Name & Description & Initial values & Initial values  \\
 &  &  &  ODE solution 1 Fig.\ \ref{sifig:cellcycle} & ODE solution 2 Fig.\ \ref{sifig:cellcycle} \\
\hline
1 & $x_1$ & Cdc2-Cyclin B & 0.9300 & 0.8400 \\
2 & $y_1$ & Wee1 & 0.9100 & 0.9900 \\
\hline
\end{tabular}
\end{table}

% Parameter Table
\begin{table}[H]
\centering
\caption{\label{tab:cellCycleParams} Parameters used in the cell cycle network.}
\begin{tabular}{|l|c|r|l|}
\hline
Parameter & Description & Value & Unit \\
\hline
$a_1$ & Production rate of $x_1$ & 1 & $s^{-1}$ \\
$a_2$ & Production rate of $y_1$ & 1 & $s^{-1}$ \\
$b_1$ & Hill coefficient for $y_1$ regulation of $x_1$ & 200 & -- \\
$b_2$ & Hill coefficient for $x_1$ regulation of $y_1$ & 10 & -- \\
$\gamma_1$ & Hill exponent for $y_1$ regulation of $x_1$ & 4 & -- \\
$\gamma_2$ & Hill exponent for $x_1$ regulation of $y_1$ & 4 & -- \\
$K_1$ & Half-activation constant for $x_1$ regulation of $y_1$ & 2.3403 & -- \\
$K_2$ & Half-activation constant for $y_1$ regulation of $x_1$ & 1 & -- \\
$A$ & External activation strength & 1 & -- \\
\hline
\end{tabular}
\end{table}

% ODE System
\begin{table}[H]
\centering
\caption{\label{tab:cellCycleODEs} The RRE ODEs for the mean-field dynamics of the cell cycle system in the limit $\Omega \to \infty$.}
\begin{tabular}{|c l|}
\hline
$\dot{x}_1$ & $= a_1 - a_1 x_1 - \frac{b_1 x_1 \left( (A y_1)^{\gamma_1} \right)}{K_1^{\gamma_1} + (A y_1)^{\gamma_1}}$ \\[0.5em]
$\dot{y_1}$ & $= a_2 - a_2 y_1 - \frac{b_2 y_1 \left( x_1^{\gamma_2} \right)}{K_2^{\gamma_2} + x_1^{\gamma_2}}$ \\
\hline
\end{tabular}
\end{table}

% SSA Propensity Table
\begin{table}[H]
\centering
\caption{\label{tab:cellCycleSSA} cell cycle reactions and the corresponding propensity functions for SSA simulations.}
\begin{tabular}{|c|l|}
\hline
Reaction & Propensity function \\
\hline
$X_1 \xrightarrow{a_1} X_1$ & $\pi_1 = a_1 \cdot \Omega$ \\
$X_1 \xrightarrow{} \emptyset$ & $\pi_2 = Y_1$ \\
$X_2 \xrightarrow{b_1} X_2$ & $\pi_3 = \frac{b_1 Y_1 \left( (A Y_2)^{\gamma_1} \right)}{(K_1 \Omega)^{\gamma_1} + (A Y_2)^{\gamma_1}}$ \\
$X_2 \xrightarrow{} \emptyset$ & $\pi_4 = Y_2$ \\
$Y_1 \xrightarrow{a_2} Y_1$ & $\pi_5 = a_2 \cdot \Omega$ \\
$Y_1 \xrightarrow{} \emptyset$ & $\pi_6 = Y_2$ \\
$Y_2 \xrightarrow{b_2} Y_2$ & $\pi_7 = \frac{b_2 Y_2 \left( (Y_1)^{\gamma_2} \right)}{(K_2 \Omega)^{\gamma_2} + (Y_1)^{\gamma_2}}$ \\
$Y_2 \xrightarrow{} \emptyset$ & $\pi_8 = Y_3$ \\
\hline
\end{tabular}
\end{table}

\subsection{The somitogenesis switch reaction network} \label{sisec:somito}

\begin{figure}[H]
\centering
\includegraphics[scale=0.9]{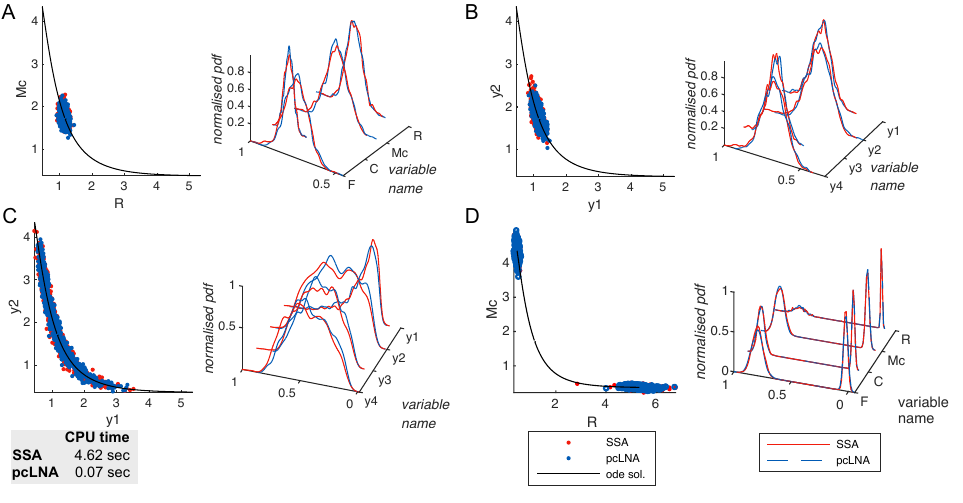}
\caption{\label{sifig:SI_devGoldbeter_omega500} 
Comparison between SSA (red) and pcLNA (blue) for the somitogenesis switch network at $\Omega = 300$. 
Panels (A)–(D) correspond to times $t=2,\ 5,\ 20,\ 99$, respectively. 
Each panel shows (left) the states recorded at time $t$ of the simulations with solutions of the corresponding RRE, and (right) the corresponding empirical probability density functions of each system variable. 
The initial conditions for the two ODE solutions are provided in Table~\ref{tab:somitoSpecies}, and the parameter values are given in Table~\ref{tab:somitoParams}.
}
\end{figure}

The system is described in \cite{Goldbeter2007}.
The tables \ref{tab:somitoSpecies}, \ref{tab:somitoParams}, \ref{tab:somitoODEs}, \ref{tab:somitoSSA} describe the species and their initial conditions, the parameters, the ODE system, and the propensity functions of the SSA simulations of this system, respectively. 
The results in Figure \ref{sifig:SI_devGoldbeter_omega500} are derived using the same method as used for Figure 3 in the main paper. 
The computation of the empirical density functions is done as described in SM section~\ref{SIsec:Fig3A}.
We used 1000 trajectories of each simulation algorithm to produce the results. 

% somitogenesis switch Tables

\begin{table}[H]
\centering
\caption{\label{tab:somitoSpecies} Species in the somitogenesis switch network and their initial concentrations.}
\begin{tabular}{ |c|c|c|c|c| }
\hline
 & Name & Description & Initial values  & Initial values  \\
 &  &  & ODE solution 1 Fig.\ \ref{sifig:SI_devGoldbeter_omega500} & ODE solution 2 Fig.\ \ref{sifig:SI_devGoldbeter_omega500}\\
\hline
1 & $R$ & Retinoic Acid (RA) & 1.1500 & 1.2000 \\
2 & $M_C$ & cyp26 mRNA & 1.7467 & 1.7067 \\
3 & $C$ & CYP26 protein & 6.2333 & 6.0900 \\
4 & $F$ & FGF8 protein & 0.4933 & 0.4800 \\
\hline
\end{tabular}
\end{table}

\begin{table}[H]
\centering
\caption{\label{tab:somitoParams} Parameters used in the somitogenesis switch network.}
\begin{tabular}{|l|c|r|l|}
\hline
Parameter & Description & Value & Unit \\
\hline
$k_{s1}$ & RA synthesis rate via RALDH2 & 1 & $s^{-1}$ \\
$\text{RALDH2}$ & RA synthetase concentration & 7.1 & -- \\
$k_{d1}$ & RA degradation via CYP26 & 1 & $s^{-1}$ \\
$k_{d5}$ & Basal RA degradation & 0 & $s^{-1}$ \\
$V_0$ & Basal transcription of $M_C$ & 0.365 & $s^{-1}$ \\
$V_{\text{sc}}$ & Activated transcription of $M_C$ & 7.1 & $s^{-1}$ \\
$K_a$ & Activation constant for $M_C$ production & 0.5 & -- \\
$n$ & Hill coefficient for FGF activation of $M_C$ & 2 & -- \\
$k_{d3}$ & mRNA degradation rate & 1 & $s^{-1}$ \\
$k_{s2}$ & Translation rate of CYP26 & 1 & $s^{-1}$ \\
$k_{d2}$ & CYP26 degradation rate & 0.28 & $s^{-1}$ \\
$k_{s3}$ & FGF8 synthesis rate & 1 & $s^{-1}$ \\
$M_0$ & Baseline activator level for FGF synthesis & 5 & -- \\
$K_I$ & Inhibition constant for RA effect on FGF & 0.5 & -- \\
$m$ & Hill coefficient for RA inhibition of FGF & 2 & -- \\
$k_{d4}$ & FGF degradation rate & 1 & $s^{-1}$ \\
$L$ & Normalization constant for spatial scaling & 50 & -- \\
$A$ & External activation strength & 1 & -- \\
\hline
\end{tabular}
\end{table}

\begin{table}[H]
\centering
\caption{\label{tab:somitoODEs} The RRE ODEs for the mean-field dynamics of the somitogenesis switch in the limit $\Omega \to \infty$.}
\begin{tabular}{|c l|}
\hline
$\dot{R}$ & $= k_{s1} \cdot \text{RALDH2} - k_{d1} R C - k_{d5} R$ \\[0.5em]
$\dot{M}_C$ & $= V_0 + \frac{V_{\text{sc}} F^n}{K_a^n + F^n} - k_{d3} M_C$ \\[0.5em]
$\dot{C}$ & $= k_{s2} M_C - k_{d2} C$ \\[0.5em]
$\dot{F}$ & $= k_{s3} \cdot \frac{M_0 A / L \cdot K_I^m}{K_I^m + R^m} - k_{d4} F$ \\
\hline
\end{tabular}
\end{table}

\begin{table}[H]
\centering
\caption{\label{tab:somitoSSA} Somitogenesis switch reactions and their corresponding propensity functions for SSA simulations.}
\begin{tabular}{|c|l|}
\hline
Reaction & Propensity function \\
\hline
$\emptyset \xrightarrow{k_{s1} \text{RALDH2}} R$ & $\pi_1 = \Omega \cdot k_{s1} \cdot \text{RALDH2}$ \\
$R + C \xrightarrow{k_{d1}} \emptyset$ & $\pi_2 = \frac{k_{d1} R C}{\Omega}$ \\
$R \xrightarrow{k_{d1}} \emptyset$ & $\pi_3 = k_{d5} R$ \\
$\emptyset \xrightarrow{V_0} M_C$ & $\pi_4 = \Omega \cdot V_0$ \\
$\emptyset \xrightarrow{V_{\text{sc}}} M_C$ & $\pi_5 = \frac{\Omega \cdot V_{\text{sc}} F^n}{(\Omega K_a)^n + F^n}$ \\
$M_C \xrightarrow{k_{d3}} \emptyset$ & $\pi_6 = k_{d3} M_C$ \\
$C \xrightarrow{k_{s2}} C$ & $\pi_7 = k_{s2} M_C$ \\
$C \xrightarrow{k_{d2}} \emptyset$ & $\pi_8 = k_{d2} C$ \\
$\emptyset \xrightarrow{k_{s3}M_0 A / L} F$ & $\pi_9 = \frac{\Omega \cdot k_{s3}  \cdot (M_0 A / L) \cdot (\Omega K_I)^m}{(\Omega K_I)^m + R^m}$ \\
$F \xrightarrow{k_{d4}} \emptyset$ & $\pi_{10} = k_{d4} F$ \\
\hline
\end{tabular}
\end{table}

\end{document}